\documentclass[fleqn,usenatbib,useAMS]{mnras}

\usepackage{graphicx}	% Including Figure files
\usepackage{amsmath}	% Advanced maths commands
\usepackage{amssymb}	% Extra maths symbols
\usepackage{multicol}        % Multi-column entries in tables
\usepackage{bm}		% Bold maths symbols, including upright Greek
\usepackage{pdflscape}	% Landscape pages
\usepackage{lscape}
\usepackage{lipsum}
\defcitealias{gon2015}{GDN15}
\defcitealias{raveDR5}{RAVE-DR5}
\usepackage[T1]{fontenc}
\usepackage{ae,aecompl}

\usepackage{newtxtext,newtxmath}
\newcommand\kms{$\mbox{km s}^{-1}$}

\title[Multiplicity among the CSGs in the MCs]{Multiplicity among the cool supergiants in the Magellanic Clouds}

\author[R. Dorda, \& L.R. Patrick] {
R. Dorda$^{\rm 1,2}$\thanks{Contact e-mail: rdorda@iac.es}, and
L. R. Patrick$^{\rm 1,2,3}$\\
$^{1}$Instituto de Astrof\'isica de Canarias, E-38205 La Laguna, Tenerife, Spain\\
$^{2}$ Universidad de La Laguna, Dpto. Astrof\'isica, E-38206 La Laguna, Tenerife, Spain\\
$^{3}$Departamento de F\'{\i}sica, Ingenier\'{\i}a de Sistemas y Teor\'{\i}a de la Se\~nal, Universidad de Alicante, E-03690 San Vicente del Raspeig, Alicante, Spain.
}

\date{Received dd mm yyyy / Accepted dd mm yyyy}

\pubyear{2020}

\begin{document}
\label{firstpage}
\pagerange{\pageref{firstpage}--\pageref{lastpage}}
\maketitle

\begin{abstract}
The characterisation of multiplicity among of high-mass stars is of fundamental importance to understand their evolution, the diversity of observed core-collapse supernovae and the formation of gravitational wave progenitor systems. Despite that, until recently, one of the final phases of massive star evolution  -- the cool supergiant phase -- has received comparatively little attention.
In this study we aim to explore the multiplicity among the cool supergiant (CSG) population in the Large and Small Magellanic Clouds (LMC and SMC, respectively). To do this we compile extensive archival radial velocity (RV) measurements for over 1\,000 CSGs from the LMC and SMC, spanning a baseline of over 40 years. By statistically correcting the RV measurements of each stellar catalogue to the Gaia DR2 reference frame we are able to effectively compare these diverse observations.
We identify 45 CSGs where RV variations cannot be explained through intrinsic variability, and are hence considered binary systems. 
We obtain a minimum binary fraction of $15\pm4\%$ for the SMC and of $14\pm5\%$ for the LMC, restricting our sample to objects with at least 6 and 5 observational epochs, respectively. Combining these results, we determine a minimum binary fraction of $15\pm3\%$ for CSGs.
These results are in good agreement with previous results which apply a correction to account for observational biases.
These results add strength to the hypothesis that the binary fraction of CSGs is significantly lower than their main-sequence counterparts.
Going forward, we stress the need for long-baseline multi-epoch spectroscopic surveys to cover the full parameter space of CSG binary systems.
\end{abstract}


\section{Introduction}    
\label{sec:introduction}

Multiplicity among high-mass stars\footnote{Those which are massive enough to end their lives as core core-collapse supernovae, which require initial masses higher than 8--10\:M$_{\odot}$ depending on metallicity~\citep{doh2015}.} has been studied observationally by many authors, and the general conclusion is that almost all of these stars are born with one or more companions.
For example \citet{san2012} estimate that 75\% of O-type stars are born with a companion, and most of these will interact at some point during their lives \citep[see also][]{san2014,kob2014}.
Indeed taking into long period systems and B-type stars \citet{moe2017} argue that 80--90\% of all high-mass stars will interact with a companion.
This interaction may have profound effects on their evolution \citep{min2013}, the nature of their subsequent supernova explosions \citep{pod1992,mar2017a,hir2020}, the formation of mergers and compact-object binaries \citep{mar2017b}, and potentially in the generation of gravitational waves \citep{bel2002}. 
Stellar mergers and mass transfer may also have an important role in shaping the observed colour-magnitude diagrams of young clusters \citep{bea2019,bri2019,wan2020}.

Evolved high-mass stars such as cool supergiants\footnote{The term red supergiant (RSG) is commonly used because most of these stars in the Galaxy are M-type. However, in lower metallicity galaxies their types are significantly earlier \citep[late-G or early-K; ][]{dor2016a}. Thus, the more general term cool supergiants (CSGs) is adopted here.} (CSGs) are typically considered as a key constraint for single-star evolutionary models \citep{eks2013}.
In addition, CSGs are the immediate progenitors of most core-collapse supernovae and consequently, compact objects. 
Given their importance to stellar feedback and gravitational wave progenitors~\citep{2020A&A...638A..39L}, understanding the multiplicity properties of this stage of evolution is vital to comprehend the observed distribution of core-collapse supernovae and gravitational wave progenitors.
Moreover, the comparison of the multiplicity in the main sequence and in the CSG phase is important to characterise the different evolutionary pathways when binary interaction is taken into account \citep{wan2020}.

Despite the key role of CSGs in these topics, their multiplicity properties have been comparatively neglected, compared with OB-type stars.
\citet{bur1983}, compiled multi-epoch radial velocities (RVs) for 181 F--M supergiants in the Milky Way (of which only 25 are M-type, and not all of them are high-mass stars) and estimated a binary fraction of 31\,--\,38\%.
\citet{neu2018,neu2019} explored the single epoch spectra of 749 CSGs from the MCs, M31 and M33 and found evidence for the presence of a B-type spectrum in 87 CSGs of their sample.
Later, \citep{neu2020} used the photometry of the LMC RSGs with a B-type component to train a kNN-algorithm. By applying this algorithm to their CSG photometric sample, they estimated a binary fraction of 19.5$^{+7.6}_{-6.7}$\%.
\citet{pat2019,pat2020} studied multi-epoch RVs of CSGs in both the 30~Doradus region of the LMC \citep{pat2019}, 14 stars, and in the cluster NGC~330 of the SMC \citep{pat2020}, 9 stars.
They obtained a binary fraction of 30$\pm10$\% in both samples, in reasonable agreement with \citet{bur1983}, but significantly lower than is found in main sequence high-mass stars.
\citet{pat2019,pat2020} argued that this difference is expected, as binary interactions would have prevented the evolution of their components toward regular CSGs in many systems.
However, their results are based on the study of a small number of CSGs and they must be tested on larger samples.

Multi-epoch RV studies that aim to identify binary motion of CSGs are challenging as the sizes of these stars (several hundreds of solar radii) impose lower limits to the orbital periods ($P_{\rm orb}$) of several hundreds--thousands of days, depending on temperature and luminosity.
Furthermore, they may exhibit intrinsic RV variations as large as $10$\:\kms{} \citep{jos2007,gra2008,kra2019}, due to their large-scale convective cells \citep{sch1975,chi2009,sto2010} that complicate the detection of orbital velocities.
In this paper we explore the largest multi-epoch sample of CSGs to date, almost 1\,000 stars from both MCs, taking advantage of the large spectroscopic catalogues of CSGs available in the literature over the last 40 years.
Although the observation frequency is low, observations are separated by years or even decades, allowing the study of long-period CSGs.

In Section~\ref{sec:sample} we present the catalogues used, their characteristics and how we combined them into a homogeneous sample.
The method for the identification of binary systems is presented in Section~\ref{sec:analysis}, and we discuss the results and their implications in Section~\ref{sec:discussion}.

\section{The Cool supergiant sample}
\label{sec:sample}

\subsection{A cool supergiant reference catalogue}

To build our sample, we searched the literature for large catalogues focused on MC CSGs.
We compile measurements from six large catalogues, described below,  because to correct RVs from each catalogue of systematic effects requires large samples (see Section~\ref{subsec:homogeneisation}

We collected all the data from each catalogue, specifically including both CSGs and other stars, to improve the homogenisation statistics, see Section \ref{subsec:homogeneisation}.
Then we cross-matched all the targets, creating a list of unique stars. 
We tagged the targets as CSGs or non-CSGs using the spectral type (SpTs) classifications provided by the catalogues.
In some catalogues, a photometric selection criteria was used, filtered by the measured RV, rather than a spectral classification.
Therefore, whenever a discrepancy was found between these catalogues and others in which spectral classification was performed, we used the identification of the latter.
In addition, we used the information available in {\sc SIMBAD} database \citep[from sources as OGLE; ][]{sos2009,sos2011} to identify Cepheids and AGB-type stars and to tag them as non-CSGs. 
The final catalogue has 1\,200 unique stars confirmed as CSGs.

\subsubsection{Prevot et al. 1985 and Mauron et al. 1987}
These two papers are part of a series of 8 articles that studied different samples of stars from the southern hemisphere with the photoelectric scanner CORAVEL.
\citet{pre1985} provide RV measurements for 404 F- to M-type supergiants in the direction of the LMC, while \citet{mau1987} did the same for 233 F- to M-type stars in the direction of the SMC (and the surrounding region).
These observations were performed over a baseline of several years, from 1981 to 1984 in the case of the LMC, and to 1985 in the case of the SMC data. Additionally, these authors provided more than one epoch for a number of targets.
Both \citet{pre1985} and \citet{mau1987} provided SpTs for their respective samples and used their RV measurements to confirm membership. The average RV errors are 1.5 and 1.3\:\kms, for the LMC and SMC respectively. 

\subsubsection{Massey \& Olsen 2003}
\citet{mas2003b} presented spectroscopic observations for a large number of red stars in the direction of both MCs; 118 towards the SMC and 167 towards the LMC.
The targets were observed during three nights (4-6 October) of 2001, where only one epoch is provided for all targets.
These authors identified CSGs (the vast majority of targets) through spectral classification and RV confirmation.
An average error of 0.3\:\kms{} is provided for all observations in this sample. 

\subsubsection{Neugent et al. 2012 and 2013}
In these two papers Neugent et al. observed large samples of candidate CSGs from the LMC (close to 2\,000 targets; 2012) and yellow supergiants from the SMC ($\sim500$ targets; 2010).
The SMC data were observed during one run of five nights in 2009 October and LMC targets were observed during one run of eight nights in January 2011.
Thus, only one epoch is provided for the stars of each galaxy.
In both studies, these authors do not provide SpTs for their targets, but determined their nature indirectly, using RV measurements to determine membership.
Each target has different a quality flag that determines the RV errors.
The average value for these errors are 2.7\:\kms{} for the SMC and 4.2\:\kms{} for the LMC.

\subsubsection{Gonz\'alez-Fern\'andez et al. 2015}
\citet[][GDN15 thereafter]{gon2015} presented multi-epoch observations between 2010 and 2013 for 544 unique stars in the SMC (315~CSGs) and 289 in the LMC (229~CSGs).
Initially, observations for both galaxies were taken in one three-night run in August 2010, and SMC targets were observed again in a two-night run in July 2011.
An additional epoch for each galaxy (this time including a large amount of candidate CSGs in addition) was observed in July 2012 and November 2013 for the SMC and LMC, respectively.
These authors identified CSGs through SpT classification, and confirmed their nature by RV measurements.
A 1--$\sigma$ average value of 1\:\kms{} is provided for all targets.

\subsection{Generic catalogues}

We cross-matched the list previously obtained from the CSGs catalogues, including both, CSGs and non-CSGs, with two large generic (i.e. not focused on CSGs) catalogues:

\subsubsection{RAVE-DR5}
\citetalias{raveDR5} observed 454 of our stars (108 CSGs), and a significant number of them over multiple epochs, between December 2004 and November 2012.
They provided individual RV measurements for each epoch and individual errors for the RV measurements, with an average value of 1.8\:\kms. 

\subsubsection{Gaia-DR2}
\label{subsub:gaia}

Gaia-DR2 \citep{gaiaDR2} provided RV measurements for 1\:436 stars from our sample, including 916 of our CSGs (92\% of the CSGs in our sample).
The current data release (DR2) provide only one RV value for each star, which is the averaged value of all the measurements obtained during the almost 2 years from July 2014, until May 2016.
The RV uncertainty of any binary may be greatly overestimated \citep{kat2019}, as the dispersion of the measures is increased by the RV variations due to orbital motion and/or intrinsic variations.
In fact, \citet{kat2019} showed that the distribution of RV uncertainties for Gaia (their Figure~10) peaks between 1 and 2\:\kms.
Above 2\:\kms{} there is a long tail composed by ``a mix of stars with insufficient signal to be processed in Gaia DR2, large amplitude variables and undetected binary or multiple systems''.

In our sample we expect both effects (intrinsic variations and orbital motions).
Thus, we may expect overestimated uncertainties for our sample.
In fact, 11\% of our CSGs have RV uncertainties in the range between 2 and 12\:\kms, with an average of 3.1\:\kms.
This will affect to our determination of what is a significant RV variation in Sect.~\ref{subsec:detection}.
In order to not overestimate the error of those CSGs with large RV variations, we truncate the uncertainties for our sample.
The value chosen is the third quartile for the distribution of RV uncertainties for Gaia: 2.08\:\kms, which delimits the beginning of the tail of over-estimated uncertainties.
This change only affects to 90 of our CSGs, which is only the upper decile.

\subsection{Homogenisation of the radial velocities}
\label{subsec:homogeneisation}

Combining RVs from different sources has the problem that there may be systematic differences between individual catalogues, as result of the instruments and methods used in each one.
Therefore, it is necessary to homogenise the RV measurements to compare the different epochs of each CSG. 
To do this, we use Gaia DR2 as reference framework.
For each catalogue (including also RAVE-DR5) and each galaxy, we calculated the RV differences between their targets and their corresponding counterparts in Gaia-DR2.

\begin{figure*}
   \centering
   \includegraphics[width=\columnwidth]{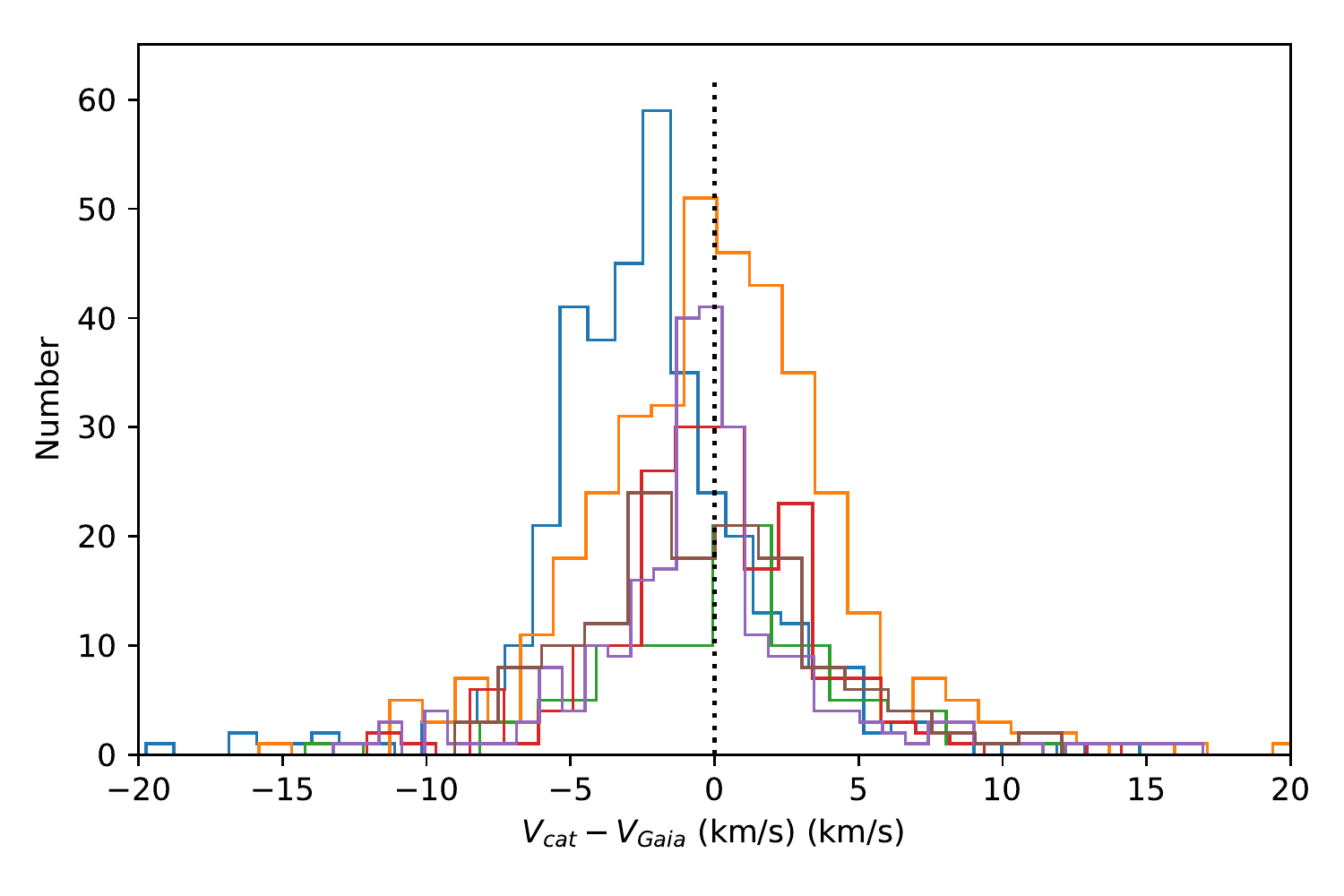}
   \includegraphics[width=\columnwidth]{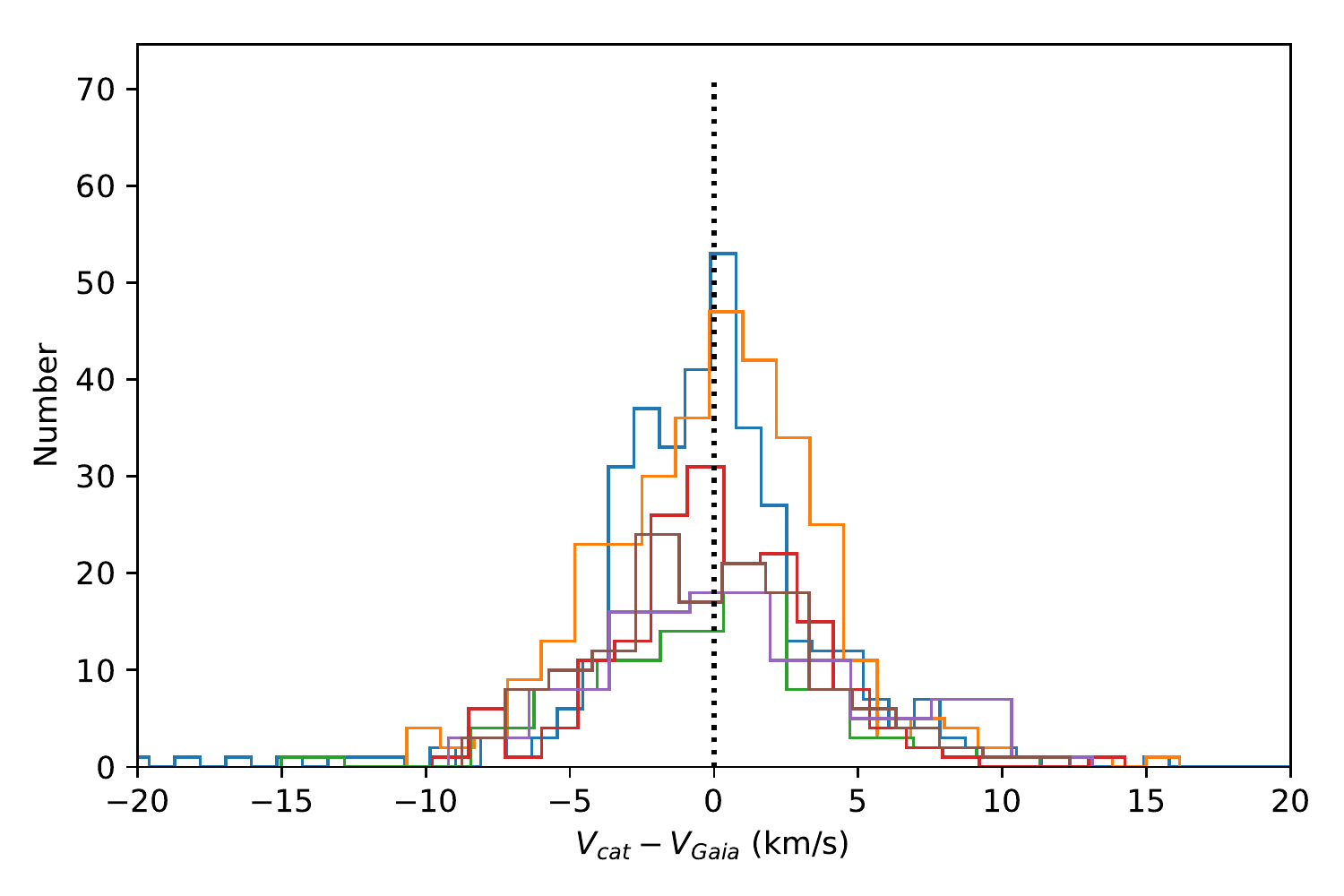} 
   \caption{
   Distribution of differences in RV between the sources of each catalogue and their counterparts in Gaia-DR2 for the LMC sample.
   The colour indicates the catalogue: orange for \citet{pre1985},  brown for \citet{mas2003b}, purple for \citetalias{raveDR5}, blue  for \citet{neu2012}, green and red for \citetalias{gon2015} data from 2010 and 2013 respectively.
   {\bf Left:} The whole catalogues (including CSGs and non-CSGs), before the correction.
   {\bf Right:} The CSGs from each catalogue, after the correction has been applied.
   }
   \label{fig:difs_LMC}
\end{figure*}

\begin{figure*}
   \centering
   \includegraphics[width=\columnwidth]{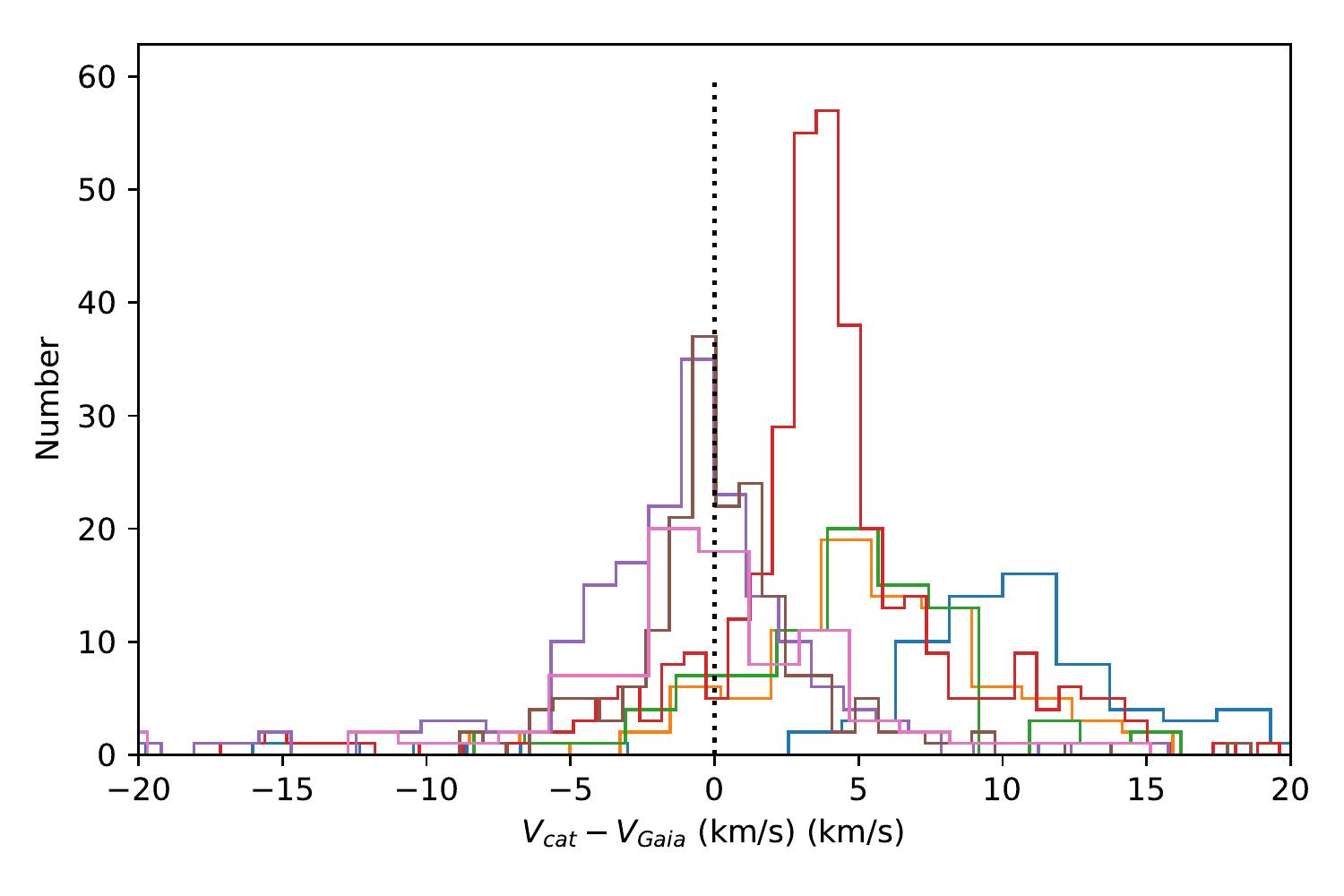}
   \includegraphics[width=\columnwidth]{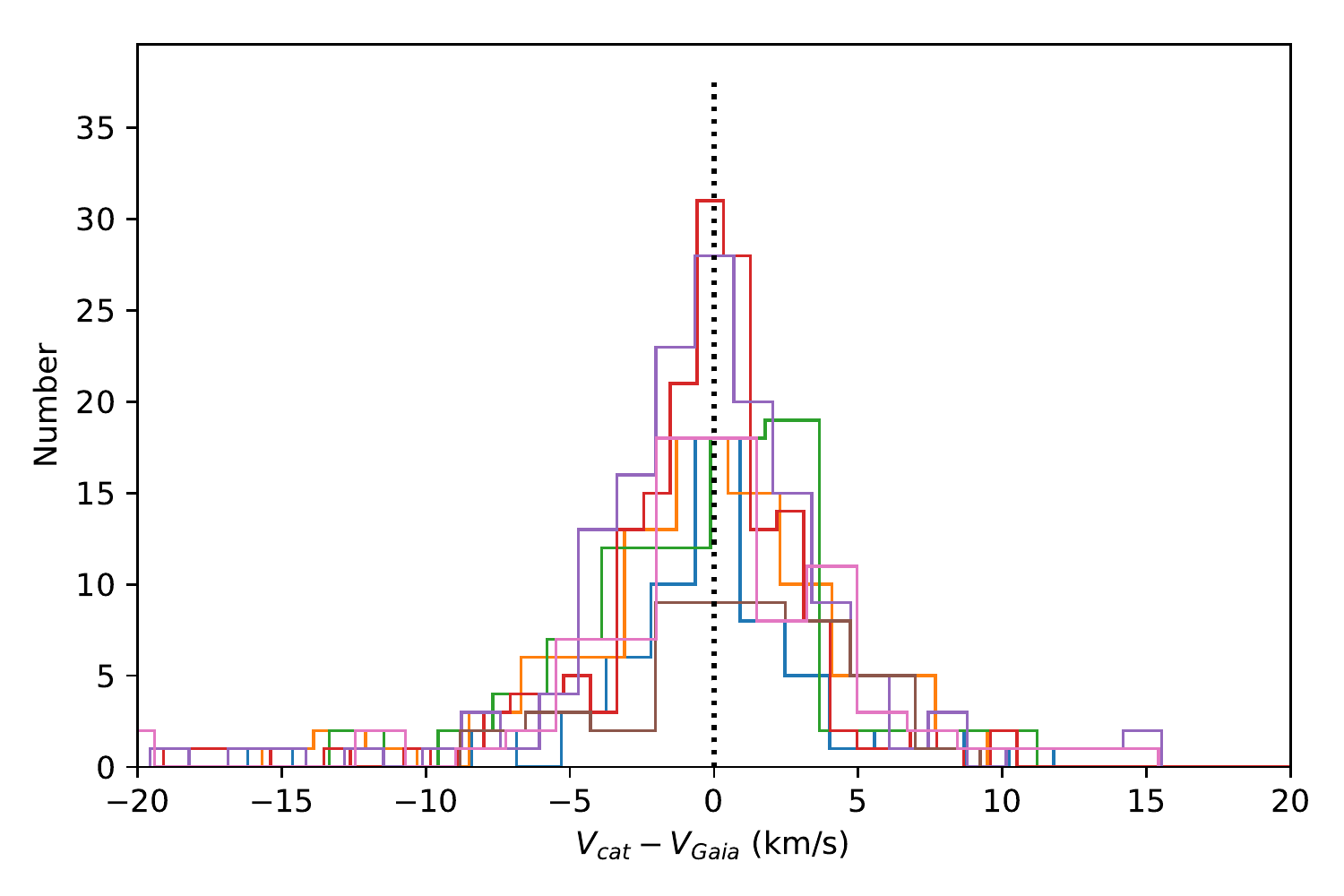} 
   \caption{Distribution of differences in RV between the sources of each catalogue and their counterparts in Gaia-DR2 for the SMC sample. 
   The colour indicates the catalogue: purple for \citet{mau1987}, pink for \citet{mas2003b}, brown for \citetalias{raveDR5}, blue for \citet{neu2010}, orange, green and red for \citetalias{gon2015} data from 2010, 2011, and 2012 respectively.
   {\bf Left:} The whole catalogues (including CSGs and non-CSGs), before the correction.
   {\bf Right:} The CSGs from each catalogue, after the correction has been applied.
   }
   \label{fig:difs_SMC}
\end{figure*}

As explained in Sect.~\ref{sec:analysis}, CSG have intrinsic RV variations.
However, as these variations are random and we use a large sample, the average offset should be zero in absence of systematic differences.
This statistical method works better with a larger sample-size, as the error of the median decreases with $1/\sqrt{n}$.
As a consequence, we decided to include the non-CSGs in each catalogue to calculate this correction.
Most of these stars are evolved intermediate-mass stars from the MCs (AGBs or carbon stars), or foreground red stars (Galactic dwarfs or halo giants).
We show the differences of the stars from each catalogue with respect to their Gaia counterparts in the left panels of Figs.~\ref{fig:difs_LMC} and~\ref{fig:difs_SMC}.

The median value of each distribution of RV differences indicate the systematic deviation of that catalogue with respect to Gaia.
Thus, we applied the median value of each distribution as a correction factor for all the sources in the corresponding catalogue and galaxy. These values are provided in Table~\ref{table:corrections}.

Once the correction is applied, we remove all the non-CSGs from our samples, as well as those CSGs with only one epoch.
We show the corrected samples of CSGs in the right panels of Figs.~\ref{fig:difs_LMC} and~\ref{fig:difs_SMC}.
The result is a multi-epoch-RV sample of almost $1\,000$ stars confirmed as CSGs, 303 in the SMC and 693 in the LMC.

\begin{table*}
\caption{Correction factor for the RVs of each catalogue, with its corresponding error (calculated using $1.253\cdot\sigma/\sqrt{n}$, as the correction factor is a median). We also show the sample size in each case (stars of each catalogue in common with Gaia-DR2).}
\label{table:corrections}
\centering
\begin{tabular}{c c | c c c }
\hline\hline
\noalign{\smallskip}
&&Correction&Median standard& Number of \\
Catalogue&Galaxy&(\kms)&error (\kms)&stars used\\
\noalign{\smallskip}
\hline
\noalign{\smallskip}
\citet{mau1987} & SMC &-0.8&$\pm0.5$&186\\
\citet{mas2003b} & SMC &-0.3&$\pm0.7$&88\\
\citetalias{raveDR5} & SMC &-0.0&$\pm0.3$&188\\
\citet{neu2010} & SMC &10.1&$\pm1.8$&77\\
\citetalias{gon2015} 2010 survey& SMC &5.4&$\pm0.6$&90\\
\citetalias{gon2015} 2011 survey& SMC &5.0&$\pm0.6$&87\\
\citetalias{gon2015} 2012 survey& SMC &3.9&$\pm0.4$&369\\
\noalign{\smallskip}
\hline
\noalign{\smallskip}
\citet{pre1985} & LMC &0.1&$\pm0.3$&375\\
\citet{mas2003b} & LMC &-0.3&$\pm0.4$&137\\
\citetalias{raveDR5} & LMC &-0.5&$\pm0.3$&248\\
\citet{neu2012} & LMC &-2.4&$\pm0.7$&383\\
\citetalias{gon2015} 2010 survey& LMC &0.8&$\pm0.6$&77\\
\citetalias{gon2015} 2013 survey& LMC &-0.1&$\pm0.3$&181\\
\noalign{\smallskip}
\hline
\end{tabular}
\end{table*}

\section{Analysis}
\label{sec:analysis}

In this section, we analyse the constructed sample with the aim of detecting binary motion.
To do this, we used two complementary methods to maximise the information obtained from our data.
Firstly, we study the largest RV variations of each star, looking for those that can unambiguously be identified as caused by binary motions and not by intrinsic variations.
Secondly, exploiting our long-baseline observations, we analyse radial velocity curves (RVCs) for out targets with the largest number of epochs, where we aim to identify RV variability that can not be explained by intrinsic variability.
Finally, we also cross-matched our catalogue with previous works where CSGs with spectral features suggesting they have an OB companion were reported.

\subsection{Dynamical binary detection}
\label{subsec:detection}

To identify significant RV variation as a result of binary motion we use similar criteria adopted in several studies~\citep{san2013,dun2015,pat2019}, which aimed at identifying binary motion in high-mass stars at different evolutionary stages.
These authors use a two-staged approach in which both criteria must be simultaneously satisfied:
the first (equation~\ref{eq:dif}) determines the minimum change in RV that can be considered as a result of binarity;
the second (equation~\ref{eq:sigma}) assesses the significance (at the 3--$\sigma$ level) of such a variation. 
Mathematically:

\begin{equation} 
|v_{i}-v_{j}|>\Delta V_{\rm lim}
\label{eq:dif}
\end{equation}

\begin{equation} 
\frac{|v_{i}-v_{j}|}{\sqrt{\sigma^2_{i}+\sigma^2_{j}}} > 3
\label{eq:sigma}
\end{equation}

\begin{figure}
   \centering
   \includegraphics[width=0.5\textwidth]{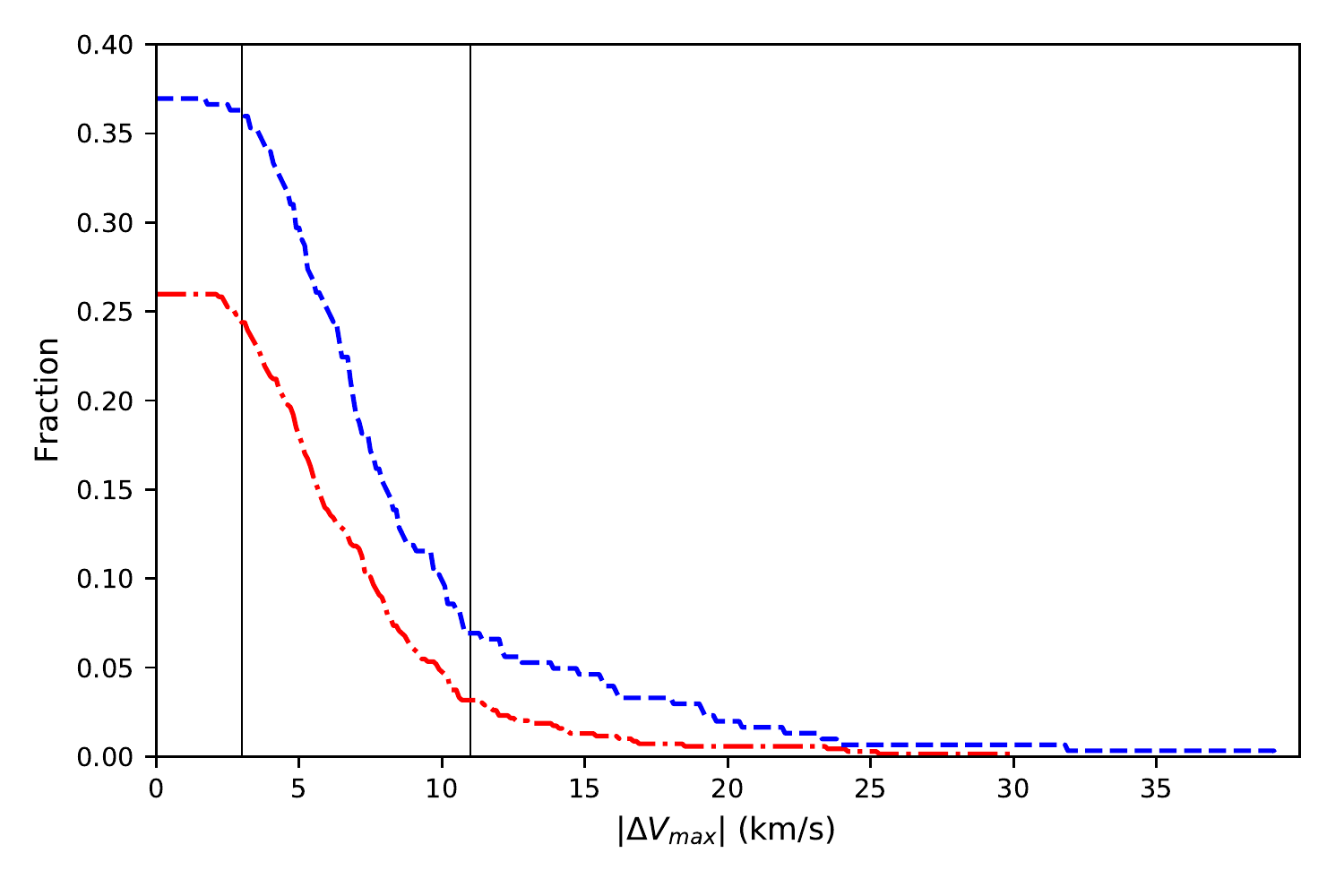}
   \caption{Cumulative distribution of the largest significant (eq.~\ref{eq:sigma})} |$\Delta V_{\rm max}$| of each CSG. 
   Red-dashed-dotted line is for the LMC sample, whereas blue-dashed line is for the SMC sample. 
   These distributions include only those CSGs with at least one significant (according to eq.~\ref{eq:sigma}) value of $|\Delta V|$, but the fraction is calculated over the total number of CSG with multi-epoch data from the corresponding galaxy.
   The black vertical lines mark the limits at $11$\:\kms, cutting the lines at $0.07$ (SMC, 21~CSGs) and $0.03$ (LMC, 22~CSGs), and at $\sim3$\:\kms{} (see text for the details).

   \label{fig:distrib}
\end{figure}

Where $v_{i}$ and $v_{j}$ are the RVs measured in any two epochs, and $\sigma^2_{i}$ and $\sigma^2_{j}$ are their respective uncertainties. $\Delta$V$_{\rm lim}$ is the limit imposed on the absolute velocity difference of any two measurements. By solely applying criterion \ref{eq:sigma}, we find that 37\% (112) and 26\% (180) of CSGs have significant RV variation in the SMC and LMC, respectively.
The results are displayed in Figure~\ref{fig:distrib}, which shows the fraction of systems that simultaneously meet both variability criteria as a function of $\Delta$V$_{\rm lim}$.

To determine a valid $\Delta$V$_{\rm lim}$, intrinsic RV variations and uncorrected instrumental effects must be considered.
It is well known that CSGs present intrinsic RV variations, due to the large convective cells that appear in their atmospheres \citep{sch1975,chi2009,sto2010}.
The observational studies \citep{jos2007,gra2008,sto2010,kra2019} show that RSGs can exhibit intrinsic variations as large as 10\:\kms.
This value can be considered as an upper limit, because larger variations are expected for larger CSGs \citep{sto2010} and the amplitude varies irregularly between cycles \citep[e.g. see][]{kra2019}.
Thus, typical RV variations for most CSGs are significantly below that limit.
In addition, our sample consists largely of G- and K-type stars (smaller in average than the M supergiants from the Galaxy for which the upper limit of 10\:\kms{} was measured).
In consequence we expect the average $\Delta$V$_{\rm lim}$ to be much smaller than this.

Despite the fact that we have no more than eight epochs for any target, we have the advantage of the size of our sample to study the multiplicity of CSGs statistically.
We examine the distribution of the RV variability of our sample (Figure~\ref{fig:distrib}) and find three distinct regions:
\begin{enumerate}
    \item Below $\sim3$\:\kms{} the distribution is truncated because most RV variations below this value are not significant according to eq.~\ref{eq:sigma}.
    \item Between 3--11\:\kms{} the slope is approximately constant, with only a small number of CSGs showing variations larger than that. This region is dominated by three factors: intrinsic variations, binary motions and uncorrected errors. While binary systems are only a fraction of the population, whose identification also depends strongly on the system inclination, intrinsic variations are expected for all CSGs. Also, any individual error not corrected by our statistical method is expected to be small (given the systematic errors found and corrected in Table~\ref{table:corrections}), and thus their effects will have much more weight here than among those few CSGs with RV variations above 11\:\kms.
    \item Above 11\:\kms{} the flattening of the curve indicates that it is no longer affected by the ubiquitous intrinsic variations. Therefore, the only explanation for these variations is orbital motion.
\end{enumerate}

By adopting $\Delta V_{\rm lim}=11$\:\kms{} in eq.~\ref{eq:dif}, we get a reliable sample of binaries.
With this limit, we find that 21~CSGs from the SMC ($6.9\pm1.4\%$\footnote{These fractions can be considered as a binomial distribution. Thus, their uncertainties can be calculated as: $\sqrt{F(1-F)/N_{\rm s}}$ where $N_{\rm s}$ is the size of the sample and $F$ is the fraction of positives in the sample ($N_{\rm p}/N_{\rm s}$).} of the total sample), and 22 ($3.2\pm0.7\%$) of the CSGs from the LMC can be considered trustworthy binaries.

\subsection{Detection through RV curves}
\label{subsec:curves}

For most of our sample, we have only two or three epochs available (see Table~\ref{table:sampling}). 
However, a small fraction of the sample present RVCs whose behaviour can be analysed.
This analysis is based on two factors: the RV variation (i.e. if the variability resembles orbital motion) and the time-scale of such variation.
Our objective is to search for RV measurements that can not be explained by intrinsic variations, but by orbital motions only.
We note that no attempt is made to estimate orbital parameters, owing to limitation in the observational data.

All RVCs with more than 4 epochs are analysed and for those that delineate a regular curve, we estimate a rough limit for the lowest period possible.
In this sense we can, in principle, distinguish between intrinsic variations and orbital motion.
The intrinsic variability of CSGs have typical periodicities on two different scales:
(i) main periods are in the range between few hundred to one thousand days and (ii) longer secondary periods up to 5\,000\:d \citep[13.7\:a; ][]{kis2006,cha2019}.
Although these periods were derived from photometric information, the extensive following of the RSG $\mu$~Cep shows that its light and RV curves are linked: they have the same periodicity with a phase shift \citep{kis2006,jor2016,kra2019}.
Thus, these periods can be compared with those derived from our RVCs.

The minimum orbital period ($P_{\rm min}$) for a CSG is limited by its stellar radius (see Appendix~\ref{appen:orbit}), as any orbit smaller than that would lead to a merger or to a mass transfer that would interrupt the normal CSG evolution in most cases \citep{eld2008}.
We calculate $P_{\rm min}$ for CSGs of different initial masses under the most optimistic circumstances (zero eccentricity), for different mass ratios ($q$).
We detail the calculations in Appendix~\ref{appen:orbit}.
We find that for any mass in the range between 9 and 25\:M$_{\odot}$, $P_{\rm min}$ is below 14\:a.
Although not all CSGs present a long secondary period, and only 14\% \citep[33 out of the 227 LMC CSGs studied by ][]{cha2019} have periods longer than 3\,000\:d, we use 5\,000\:d as a limit, to increase the reliability given our sparse sampling.
Thus, we discard RVCs whose shapes suggest a period below 14\:a.

Eight targets were initially selected with potential periods above 14\:a. 
Most of the RVCs selected exhibit a linear trend over a decade or more, suggesting periods of several decades. 
However, given the sparse sampling of the RVCs, such trends can be caused by the combination of intrinsic variations of shorter periods over larger baselines.
To evaluate this effect, we performed a Monte Carlo simulation where for each star we simulated $10^6$ random sinusoidal RVCs, with phases and periods determined randomly as follows.
For those CSGs in \citet{cha2019} with a known period for their intrinsic-variability, we used random periods within $\pm10\%$ of the nominal period.
For the rest, we selected randomly intrinsic-variability periods in the range 160 to 2\,000\:d. 
These period limits are chosen based on the distribution of measured periods in \citet{cha2019}.
For each simulated RVC we extract RV measurements at the same epochs as the target stars taking into account its corresponding uncertainty\footnote{We randomly calculated the RV value by drawing it from a Gaussian distribution centred in the obtained RV value, and with a sigma equal to the corresponding uncertainty.}, and for the obtained data we calculated the absolute value of the Pearson's correlation coefficient.
We compared these coefficients with those obtained from the observational data. 
We calculate the probability of having a random alignment with a correlation coefficient equal or higher than the observed one.
To consider that the correlation from the RVC observed is not product of a random alignment of data points caused by intrinsic variations, we require that the probability calculated is lower than a certain threshold. 
We chose a significance level of 0.27\%  (3--$\sigma$ for an equivalent Gaussian distribution) for these comparisons, as it is equivalent to the significance level used in the method explained in the previous subsection.
In other words, we accepted as non-random those RVCs that have a probability of change alignment of lower than $0.27\%$.

Using this methodology, we find two CSGs (both from the SMC) whose RVCs cannot be explained via a combination of random fluctuations and periodic intrinsic variability with typical periods in the range 160\,--\,2000d.
We therefore assume these trends to result from genuine orbital motion.
We provide the discussion for each individual case and we show the corresponding RVC in Appendix~\ref{appen:curves}.

\subsection{CSGs with blue-star spectral features}
\label{subsec:counterparts}

A number of CSGs in the MCs were reported in previous works as having spectral features of early-type components (blue-star spectral features; BSSFs).
Considering these targets as confirmed binaries is risky.
The probability of random alignment with a blue star is not negligible, particularly in the case of those located in crowded star-forming regions.
However, they can be considered as strong candidates.
In this section we analyse the cross-match of such CSGs with our own catalogue, which is also summarised in Table~\ref{table:crossmatch}.

\citetalias{gon2015} reported 12~CSGs in whose spectra they found also the spectrum of a OB-type star at their bluest wavelengths. 
Also, these authors reported 12~CSGs with Balmer lines in emission.
However, these lines may be caused by nebular emission. 
Thus, we checked in detail these spectra, as they are available in \citet{dor2018b}.
We have tentatively confirmed the composite nature of 4 of these stars (identified by its catalogue name), obtaining a total of 16 CSGs with BSSFs:

\noindent
\textbf{[M2002] 23463}: Emission is observed in H$\beta$, H$\gamma$ and H$\delta$ in the 2010 and 2012 data, with evidence for a slight blue asymmetry to H$\beta$. 
Furthermore there is a hint of an underlying absorption component in H$\delta$ which may suggest the presence of a B or Be star as a companion. This star is also the list presented of RSG+B-type stars for the SMC presented in \citet{neu2019}.
Unfortunately, we can not confirm for sure if it is the same Be star as they identified, because they did not provide the ID of that star.

\noindent
\textbf{YSG010}: This star has strong narrow Balmer lines in emission that may be nebular in origin. However there is clear evidence for a blue continuum in the spectrum, and we have positively identified He\,{\sc i} emission lines at 4009 and 4026 \AA{} for example, implying the secondary is a B-type star. This system is discussed further in Sect.~\ref{subsec:triple}.

\noindent
\textbf{[M2002] 55355:} Both H$\gamma$ and H$\delta$ are visible as absorption lines.
There are no clear signs of emission.

\noindent
RM2-093: The spectrum is peculiar having many emission lines in addition to the Balmer series. 
A number of these are coincident with the positions of allowed Fe\,{\sc ii} lines and may indicate that the companion is an emission line B-type star. Further data is necessary to confirm the nature of this system.

\citet{neu2019} reported 24~CSGs from the MCs with spectral features from an OB-type star (Balmer's lines too strong for a CSG).
Two of these stars are in common with the list from \citetalias{gon2015} (both of them with significant RV variations below 11\:\kms).

In total, from the 38~CSGs with BSSFs reported in literature, we have at least two epochs for 30 of them, among which 11 have no significant RV variations (according to eq.~\ref{eq:sigma}).
This is not unexpected, as 9 of them have only two or three epochs available.
The other two may be explained by high inclinations or long orbital periods, as none of them seems to be located close to any potential contaminant.

Among the 19~CSGs with significant RV variations, 10 have RV variations between 4.6 and 10.2\:\kms, and their RVCs do not exhibit clearly a binary behaviour. 
There are also two CSGs which are located in clusters, and thus, their BSSFs are possibly caused by contamination from a unrelated blue star.
The rest (seven CSGs) are considered binaries by the dynamical criterion, and we checked that none of them are located close to any potential contaminant.
Therefore, the BSSFs observed in these CSGs belong to their companions, which must to be relatively-luminous OB-type stars.

\begin{table*}
\caption{CSGs in our sample that were reported to have BSSFs in previous works. The fractions are calculated respect the number of stars observed at least in two epochs, but without taking into account any CSG located in a cluster, as their BSSFs may be caused by contamination.
}
\label{table:crossmatch}
\centering
\begin{tabular}{c c c c c c c}
\hline\hline
\noalign{\smallskip}
&CSGs with&Observed at least&Signficiant&In&Identified as binary\\
Origin&BSSFs&in 2 epochs& by eq.~\ref{eq:sigma}&Clusters&in this work\\
& \# & \# & \# & \# & \# (\%)\\
\noalign{\smallskip}
\hline
\noalign{\smallskip}
From \citetalias{gon2015}& 16 & 14 & 10 & 2 & 4 (33)\\
From \citet{neu2019}& 24 & 18 & 11 & 0 & 3 (17)\\
\noalign{\smallskip}
\hline
\noalign{\smallskip}
Total (unique CSGs$^a$) & 38 & 30 & 19 & 2 & 7 (25)\\
\noalign{\smallskip}
\hline
\multicolumn{5}{l}{\footnotesize $^a$ There are two CSGs in common between these two catalogues}\\
\end{tabular}
\end{table*}

\section{Discussion}    
\label{sec:discussion}

We have found 45 binary systems among the sample of CSGs studied (see Table~\ref{table:results}). Moreover, 41 of them, were previously unknown. In this section we explore the implications of this finding.

\subsection{A lower limit for the binary fraction}
\label{subsec:lowerlimit}

The binary fraction (BF) is the fraction of stars from a given population (CSGs in this case) in a binary (or multiple) system. 
In this work, we find that 7.6\% of the CSGs studied in the SMC and 3.2\% in the LMC, are reliable binaries (see table~\ref{table:results} for the list of binaries found).
These values are significantly lower than those obtained in previous works, but it is not unexpected.
Our method is affected by a number of biases, such as inclination, sampling, orbital parameters, etc. 
Previous works \citep{pat2019,pat2020} corrected their samples of observational biases using Monte Carlo simulations, estimating the intrinsic BF from their observed fraction.
However, to perform such corrections an orbital-parameter distribution, must be assumed.
Observationally, the orbital parameter distribution for CSGs is not well constrained. Because of this, \citep{pat2019,pat2020} extrapolated the orbital parameter distributions from earlier evolutionary phases (OB stars) \citep{moe2017}.
In the present study we have a sample large enough to estimate the minimum binary fraction (BF$_{\rm min}$) expected for CSGs with an acceptable statistical significance without assuming any orbital parameter distribution.
Therefore, our results can be used as independent constraints for evolutionary models for binary populations.
Also, in this way our results can be used to check, in an independent manner, the results obtained by \citet{pat2019,pat2020}.
In fact, it is encouraging that our result for the SMC ($7.6\pm1.5\%$) is compatible at 2--$\sigma$ with the observed BF (10\%) from \citet{pat2020} for the CSG in NGC~330, a cluster located in that galaxy, when the limit $\Delta V_{\rm lim}=11$\:\kms{} is used in both cases.

Our main bias is the very limited number of epochs.
As we show in Table~\ref{table:sampling}, the maximum number of epochs is 7 and 8 for the samples from the LMC and the SMC respectively, and only a few stars reach that level of sampling.
In addition, a significant fraction of both samples (58\% in the LMC and 40\% in the SMC) have only two epochs.
As can be deduced from Table~\ref{table:sampling}, to find more binaries (by any of the methods used here) is more likely when more epochs are available, but our sample is clearly dominated by those with a very low number of epochs.
Fortunately, we can minimise the main bias in our sample, without need of theoretical assumptions or simulations, by restricting the CSG sample that we consider to those objects with a given minimum number of epochs.
However, the drawback in this scenario is that as we increase the minimum number of epochs, fewer stars are included in the minimum BF calculation, and thus, the uncertainties (see Sect.~\ref{subsec:detection}) will be larger.
Given the fractions found previously, we decided to assume a maximum uncertainty for the percentage of 5.
This condition is fulfilled in the LMC sample using those CSGs observed in five or more epochs (51 stars), while in the SMC is fulfilled using those with six or more epochs (72 stars).
Under these limits, we found 11~binary CSGs in the SMC and 7 in the LMC, which represent $15\pm4\%$ and $14\pm5\%$, respectively, of the CSGs observed at least the number of epochs selected in each case.

Finally, we considered if there is any reason to not to combine the results from both galaxies.
The differences in typical $T_{\textrm{eff}}$ \citep[$\sim300$\:K; ][]{tab2018} between the SMC and the LMC, affect to the average sizes of the stars from both galaxies. 
However, these differences are small and, therefore, the $P_{\rm min}$ for the CSGs from both galaxies is very similar (see Fig.~\ref{fig:P_min}).
As we can expect systems in a very wide range of periods, this small difference seems unlikely to affect in a noticeable way the abundance of binaries.
In the absence of a strong theoretical argument against, together with the fact that the BF$_{\rm min}$ calculated for both galaxies from the subsamples selected are statistically indistinguishable, makes us to think that we can use the BF of both MCs combined (15$\pm3$\%), as global BF$_{\rm min}$ for CSGs.
This value agrees well with the bias-corrected results obtained by \citet{pat2019,pat2020} and \citet{neu2020} for samples from both galaxies.

\begin{table*}
\caption{CSGs split by number of RV epochs, and binaries detected among them (number and percentage).
}
\label{table:sampling}
\centering
\begin{tabular}{c | c | c c | c | c c}
\hline\hline
\noalign{\smallskip}
 & \multicolumn{3}{c|}{LMC Sample} & \multicolumn{3}{c}{SMC Sample}\\
Number of & CSGs & Binaries & Accumulated ($\geq$\# of epochs) & CSGs &  Binaries & Accumulated ($\geq$\# of epochs)\\
epochs & \# (\%$^b$) & \# (\%) & \# (\%) & \# (\%$^b$) & \# (\%)& \# (\%)\\
\noalign{\smallskip}
\hline
\noalign{\smallskip}
2 & 404 (58) & 5 (1.2)& 22 (3) & 122 (40) & 2 (1.6) & 23 (8) \\
3 & 173 (25) & 7 (4)& 17 (6) & 63 (21) & 6 (10) & 21 (12) \\
4 & 65 (9) & 3 (5)& 10 (9) & 23 (8) & 2 (9) & 15 (13)\\
5 & 32 (5) & 4 (12)& 7 (14) & 23 (8) & 2$^a$ (9) & 13 (14) \\
6 & 16 (2.3) & 2 (13)& 3 (16) & 54 (18) & 8$^a$ (15)& 11 (15) \\
7 & 3 (0.4) & 1 (33) & 1 (33) & 16 (5) & 3 (19) & 3 (17) \\
8 & -- & -- & -- & 2 (0.7) & 0 (0) & 0 (0) \\
\noalign{\smallskip}
\hline
\noalign{\smallskip}
All & 693 & 22 (3.2) & -- & 303 & 23 (7.6) & -- \\
\noalign{\smallskip}
\hline
\multicolumn{7}{l}{\footnotesize $^a$ We have taken into account here one extra binary found through the RVC method (see Sect.~\ref{subsec:curves})}\\
\multicolumn{7}{l}{\footnotesize $^b$ Calculated over the total number of CSGs}\\
\end{tabular}
\end{table*}

\subsection{Companions}
\label{subsec:possiblecompanions}

\subsubsection{Expected orbital biases}
\label{subsub:composition}

\begin{figure}
   \centering
   \includegraphics[width=\columnwidth]{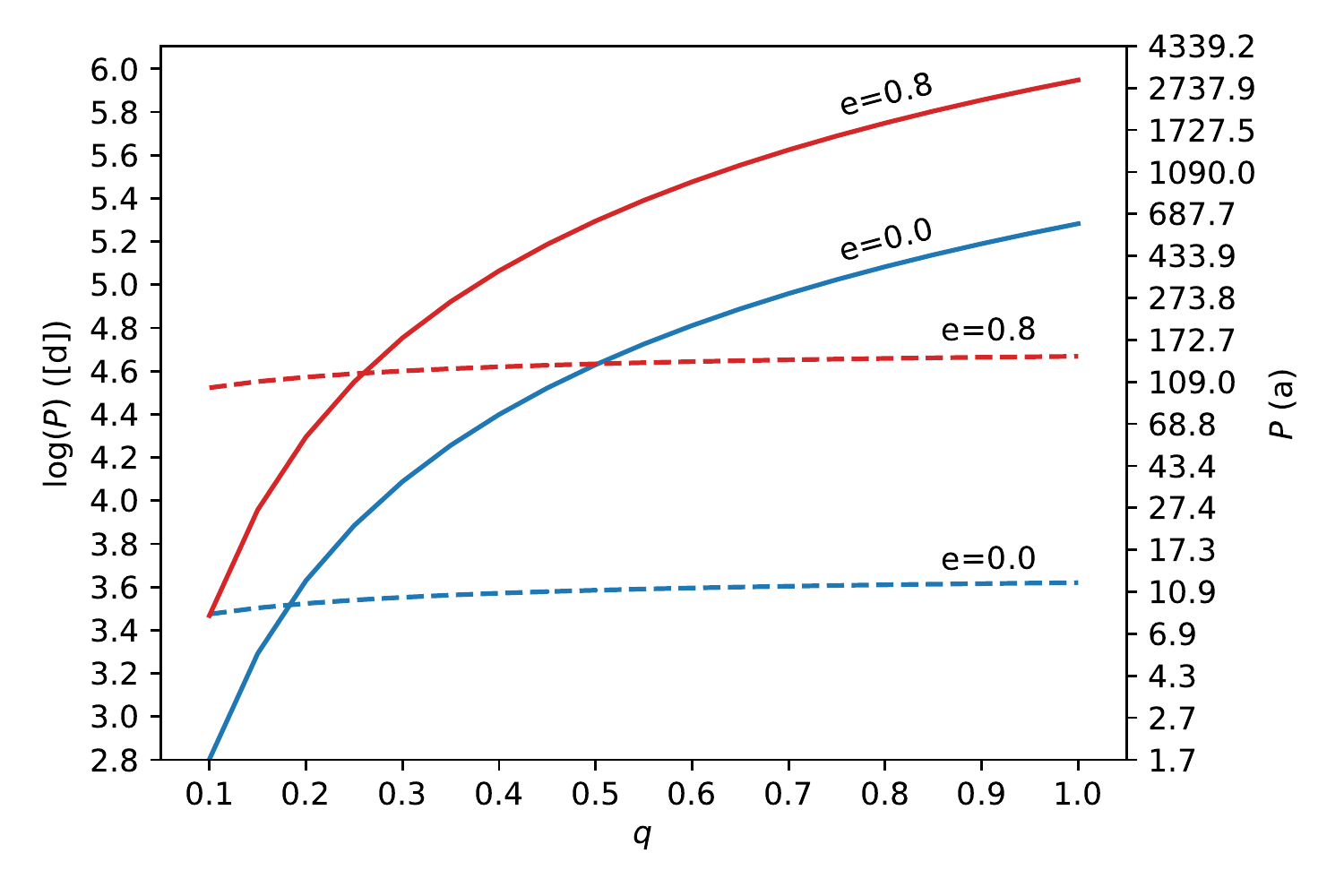}
   \includegraphics[width=\columnwidth]{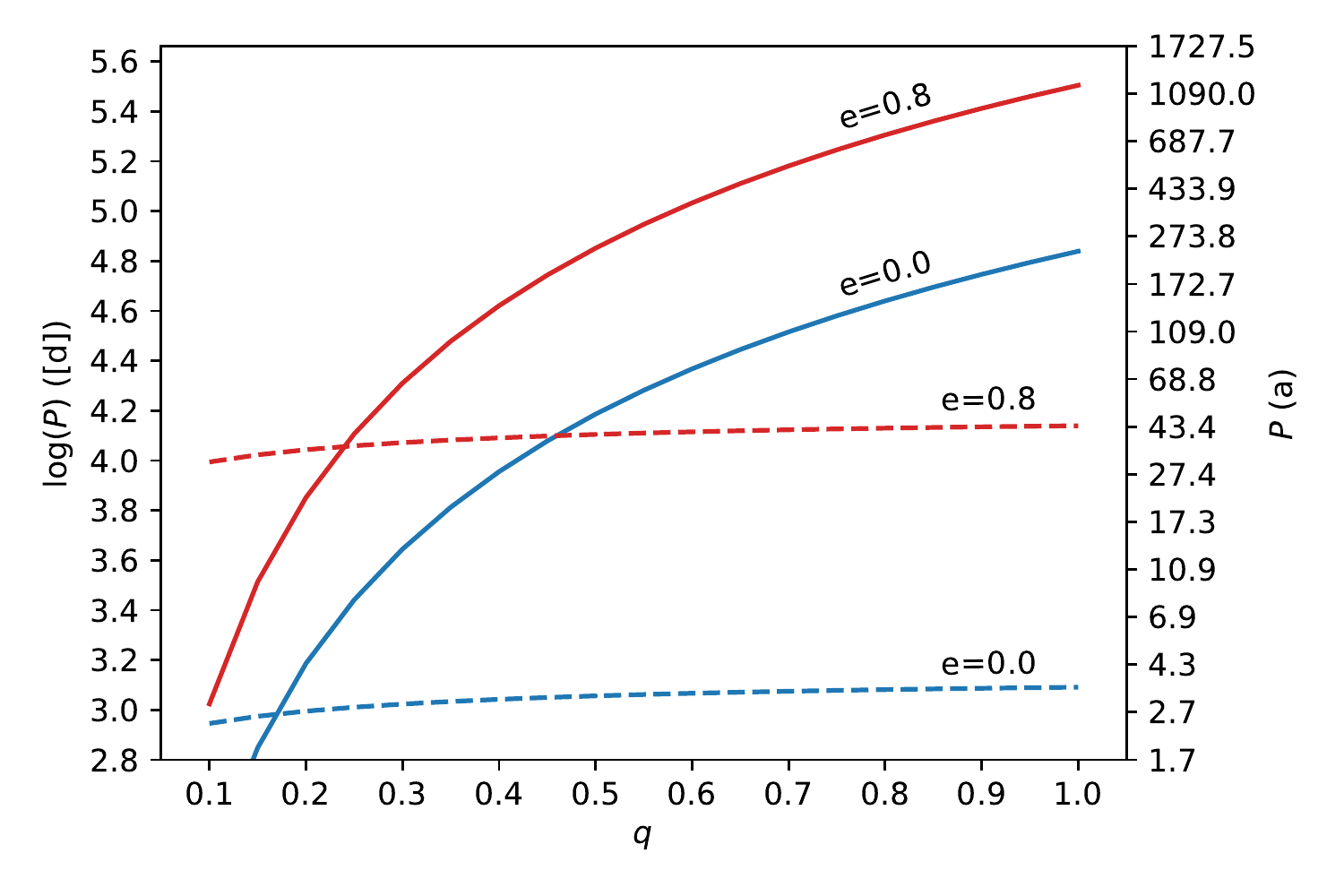}
   \caption{
   $P_{\rm orb}$\,--\,$q$ plane for CSGs from the LMC. 
   The color of the lines indicates the eccentricity used for their calculation.
   We show the results for two extreme eccentricities, 0.0 (in blue) and 0.8 (in red).
   Solid lines indicate the limit imposed by $K_{1}^{\rm min}=6.8$\:\kms{} in the $P$\,--\,$q$ plane: any binary CSG detected by our method} should be at the right of these lines.
   Dashed lines indicate the minimum periods imposed by the size and the eccentricity, for given a mass (the lower panel is for $9$\:M$_{\odot}$, whereas upper panel is for $25$\:M$_{\odot}$).
   For the SMC, the Figure is equivalent, but with minimum periods $\sim0.1\,$dex lower.
   \label{fig:param_space}
\end{figure}

A detailed characterisation of each binary system will be performed in a future paper.
Here we analyse the sample statistically, in order to understand its biases.
The dynamical binary detection method only detects systems with a RV variation larger than 11\:\kms, which limits the detectable semi-amplitude velocity ($K_{1}$) of the systems, and in turn limits the periods and mass ratios detectable by this method.
We note that this limit on $K_{1}$ does not affect binaries found through the RVC method alone.
However, such stars represent only $\sim4\%$ of binaries found and as a consequence they do not significantly affect the following analysis.

In the most optimistic scenario the measured RV variation is equal to the peak-to-peak velocity, which is twice $K_{1}$ by definition.
However, given our limited sampling (less than 10~epochs per target), the $|\Delta V_{\rm max}|$ measured in most cases is not a reliable estimate of the true binary peak-to-peak velocity (Sim\'on-D\'{\i}az et al., submitted).
In most cases the peak-to-peak velocity will be greater than the measured $|\Delta V_{\rm max}|$.

To quantify this effect, we performed a Monte Carlo simulation where $10^6$ synthetic RVCs were generated (with an unit amplitude for simplicity, and thus a peak-to-peak value of 2).
We then took a number of RV measurements ($n$) from each curve, where $n$ is randomly determined for each curve using as a probability distribution the sum by rows of the columns named ``Binaries'' in Table~\ref{table:sampling}.
By calculating the differences between the values obtained from each curve, we obtain $|\Delta V_{\rm max}|$.
Putting together the $|\Delta V_{\rm max}|$ obtained for every synthetic RVC, we obtain a distribution of values between 0 and 2 (the minimum and maximum values possible, respectively, for the $|\Delta V_{\rm max}|$ measured in the synthetic RVCs.
Despite the most probable value in the distribution is 2, the median value is only 1.62.
This simple experiment confirms that the measured $|\Delta V_{\rm max}|$ is, in general, not a reliable estimate of the peak-to-peak velocity of our stars.
In consequence, if we consider that the peak-to-peak amplitudes in our sample can be typically as low as the limit $|\Delta V_{\rm max}|=11$\:\kms, we would be too optimistic. 
Statistically, using as lower limit $|\Delta V_{\rm max}=11$\:\kms{} for binary identification implies that we are imposing a limit to the typical peak-to-peak amplitude significantly higher than 11\:\kms.
We can estimate a more realistic peak-to-peak value using the ratio between the peak-to-peak amplitude in the synthetic RVCs and the median value obtained: 1.62/2=0.81. 
Thus, we should expect that in median the 11\:\kms{} limit is equivalent to a limit in the peak-to-peak velocity corresponding of $11/0.81=13.6$\:\kms.
Due to that, the expected lower limit value for $K_{1}$ in the of observed binary systems is
$6.8$\:\kms.
Take into account that this limit is under the assumption of $i=90$\textdegree{}. 
However, an inclination $i<90$\textdegree{} implies that the real peak-to-peak velocity is even greater than the $|\Delta V_{\rm max}|$ observed, and thus that $K_{1}$ is also even higher.
In consequence the inclination does not affect to the lower limit on $K_{1}$ calculated.

$K_{1}$ is related to the orbital period ($P_{\rm orb}$), the orbital eccentricity ($e$), and the mass-ratio ($q=M_{\rm companion}/M_{\rm CSG}$).
A lower limit to $K_{1}$ defines for each eccentricity an area in the $P$\,--\,$q$ diagram where systems cannot be detected by our method.
In Figure~\ref{fig:param_space} we show the limits imposed by the minimum $K_{1}$ on orbital configurations for CSGs in the LMC as solid lines.
The detectable parameter space (for a given eccentricity) is located below these solid lines.
The minimum period possible, imposed by the size of the CSGs is also shown, which,
as explained in Sect.~\ref{subsec:curves}, depends on stellar radius, mass and eccentricity (see Appendix~\ref{appen:orbit} for details).
The dependence on eccentricity is important as, for a given $q$, a higher eccentricity implies a shorter periapsis separation and the minimum separation is limited by the physical size of the CSG.

The combination of $P_{\rm min}$ and minimum semi-amplitude velocity implies a number of biases for the binary systems detected in our sample.
We can detect $q$ values down to 0.20, over the entire CSG mass range (see Figure~\ref{fig:param_space}), in the most optimistic circumstances ($e\sim0$) and down to $q\sim$0.25 at $e=0.8$.
However, low-$q$ systems can only be detected for short orbital periods (as a result of the minimum semi-amplitude velocity cut-off).
High-$q$ systems can be detected in a wider range of periods.
This results in a favorable detection-bias toward higher $q$ systems.
In addition, systems with short periods can be detected at a wider range of $q$'s.
This biases our detections toward shorter periods.
We must consider also that high-eccentricity systems are less likely to be detected by our method than those with a lower eccentricity.
A system with a $K_{1}$ large enough to be detected but also a high eccentricity will display a detectable RV difference only during a small fraction of its orbit.
Given our limited sampling, the detection probability for high-eccentricity systems is significantly lower than for low-eccentricity systems.
Therefore, we expect to be biased toward low and intermediate eccentricities.

\citet{moe2017} studied the frequency distribution for main-sequence OB-type stars (the progenitors of CSGs). 
For those with periods $>20$\:d the companion frequency peaks at $\log(P_{\rm orb})\sim3.5$\:dex~[days] (see their Fig.~37), with a strong preference for lower $q$'s ($q\sim0.2$\,--\,$0.3$), and a eccentricity distribution weighted toward larger values.
This implies that our RV limit of 11\:\kms{} misses a significant part of the binary population: those systems with a combination of a low $q$ and a long orbital period.
Combining the observational biases together with the expected distribution of parameters, we can conclude that our sample is likely dominated by companions with intermediate $q$'s (higher than 0.20), and periods around 10$^{3.5}$\:days. As those systems with mid to high eccentricity have $P_{\rm min}$ significantly above 3.5\:dex~[days], probably most of them interacted or merged. Only those few with periods high enough would have survived. Thus, we expect that systems with low eccentricities are significantly more frequent than those with high eccentricities.

The minimum $q$ expected is $\sim0.20$ regardless of the CSG mass. As most of our sample are mid-to-high luminosity CSGs \citep[see e.g.][]{gon2015,tab2018}, the lowest mass expected is roughly 10\:M$_{\odot}$.
Therefore, the companions of these stars can have masses as low as $2$\:M$_{\odot}$.
We must take into account that companions with masses lower than $\sim3$\:M$_{\odot}$ (later than B9~V) have not entered in the main sequence by the time lowest-mass CSGs explode as supernova \citep{neu2019}.
For a typical CSG of 15\:M$_{\odot}$, the lowest mass companions should have masses around 3\:M$_{\odot}$.
Therefore, we conclude that our method is able to detect any O or B companion, although late-B types can not be detected in all cases, only in low-mass CSGs. 
In addition, we expect CSGs plus compact object systems to be detectable.
Such objects are remnants of the original primary star \citep{kru2016,kle2020} of the system, while the CSG was the original secondary. 
Neutron stars have a highest theoretical mass of 2.16\:M$_{\odot}$, which is just slightly higher than the lowest mass detectable by our method.
Thus, it is very unlikely to have detected them in this study.
On the other hand, black-holes can have higher masses, and thus they can be present among the companions detected.
However, to identify the nature of the companions in the systems found, the orbital properties of the systems must be constrained in future works.

\subsubsection{Comparison with single-epoch surveys.}
\label{subsub:bias}

The method proposed by \citet{neu2018} uses the identification of BSSFs in single-epoch spectra to detect binary CSGs.
They obtained that 4\% of their MC sample present BSSFs.
This value is significantly below our estimated minimum BF of $15\pm3\%$, but we must take into account that we are comparing different methods.
To compare these results we must consider the fraction of CSGs with BSSFs in our sample.
We can not use our whole sample for this calculation, as it has not been searched entirely looking for BSSFs.
Instead, we use the sample from \citet{gon2015}, composed by 544~CSGs for which they reported 16~CSGs with BSSFs (including the two located in clusters).
We also take into account 10~CSGs in that list with BSSFs reported by \citet{neu2019}.
These 26~CSGs imply a BF of $4.6\pm0.9\%$, which is fully compatible with the results obtained by \citet{neu2019}.
In addition, our value is compatible with the bias-corrected BF (19.5$^{+7.6}_{-6.7}$\%) obtained by \citet{neu2020} for the LMC. 
All this confirms that the difference between the results is not intrinsic to the samples considered, but to the different methodologies used.

According to \citet{neu2018}, the coolest OB companion they considered (15\,000\:K, $\sim$B5~V), can be concealed easily by a relatively warm CSGs (see the yellow-supergiant spectrum in their Fig.~7).
Moreover, they found that their warmest CSGs can conceal the contribution of even the earliest-type main-sequence B-type stars.
On the contrary, as explained in Sect.~\ref{subsub:composition}, our sample seems to be dominated by companions with $q$'s slightly above 0.25.
Such companions, if present in the main sequence, are expected to have intermediate- and late-B types.
In fact, our method is theoretically able to detect companions as late as B9~V ($\gtrsim3\,$M$_{\odot}$).
However, their BSSFs can not be detected in the CSG spectra according to \citet{neu2018}.
Thus, the presence of such companions can explain the difference between the results of each method. 

We also compare our results with those obtained for M31 and M33 by \citet{neu2019}.
Among the 149 candidates these authors observed, they found 63 RSGs with BSSFs (42\%).
This value is well above the lower limit for the BF found in this work.
However, it is much higher (one order of magnitude) than the results obtained by \citet{neu2019} for the MCs (or by us in this work, as explained above), and about twice the bias-corrected value obtained by \citet{neu2020} for the LMC.
We think that this striking difference is due to three factors.
Firstly, their samples from M31 and M33 were selected using photometric criteria (in the $U-B$ and $R-I$ plane) optimised to find RSGs with OB companions.
On the contrary their MC sample belongs to another unpublished project
focused on finding RSGs, without regard of any potential OB companion.
That is also the case of the catalogues used in the present work.
Secondly, according to \citet{neu2018} warmer RSGs can conceal hotter B companions.
As CSGs in the MCs are in average warmer than those from higher-metallicity environments (as M31 and M33), they should be more effective concealing their blue companions.
Thirdly, because of the larger distances to M31 and M33
the probability of having RSGs aligned by chance with OB-type stars is significantly greater. 
M31 and M33 candidates were observed with the fiber-fed spectrograph Hetospec, whose fibers have a projected diameter of 1.5~arcsec.
At the distance of these galaxies (778\:kpc and 840\:kpc respectively), the projected diameter of a fiber covers 5.7\:pc in M31 and 6.1\:pc in M33.
On the other hand, their MC RSGs \citep[as well those from ][]{gon2015} were observed with AAOmega, whose fibers are slightly lager (2\:arcsec of diameter).
However, the LMC and the SMC are much closer (50 and 62\:kpc respectively), and thus, the projected diameter of a fiber is significantly smaller (0.5\:pc in the LMC and 0.6\:pc in the SMC).
The probability of an alignment by chance in the MC observations is non-negligible,
However, in the case of M31 and M33, given that open clusters with RSGs have typical sizes about few pcs\footnote{e.g. NGC~2345 has a core diameter of of 5.2\:pc and a halo of 28.6\:pc \citealt{alo2019}, and given the 4\:kpc to NGC~7419 \citealt{mar2013}, its diameter can be estimated in $\sim4.9$\:pc.} the probability is significant.

Despite the fact that the multi-epoch methodology is able to find faint OB-type stars or compact-object companions that can not be detected through BSSFs, it has a number of important limitations (biased periods and eccentricities; and the effect of the orbital inclination) that do not affect to the detection of BSSFs.
In fact, as shown in Table~\ref{table:crossmatch}, among the 28 CSGs with BSSFs for which we have at least 2 epochs, we only detect as binaries seven of them ($25\%$).
This percentage is significantly higher than the percentage of binaries found in the whole sample ($25\pm8\%$ against $4.5\pm0.6\%$), and than the minimum BF calculated in Sect.~\ref{subsec:lowerlimit} ($15\pm3\%$).
This suggests that detecting BSSFs, although does not prove physical connection, is an acceptable proxy for binarity in the MCs.
However, the level of contamination increases with distance, therefore we expect this technique to become more ineffective for galaxies at larger distances.
From all these results, we conclude that both techniques should be used complimentary, as no single technique is able to provide a complete picture of the binary CSG population.

\subsection{The first CSG in a triple system?}
\label{subsec:triple}

\begin{figure*}
   \centering
   \includegraphics[width=\columnwidth]{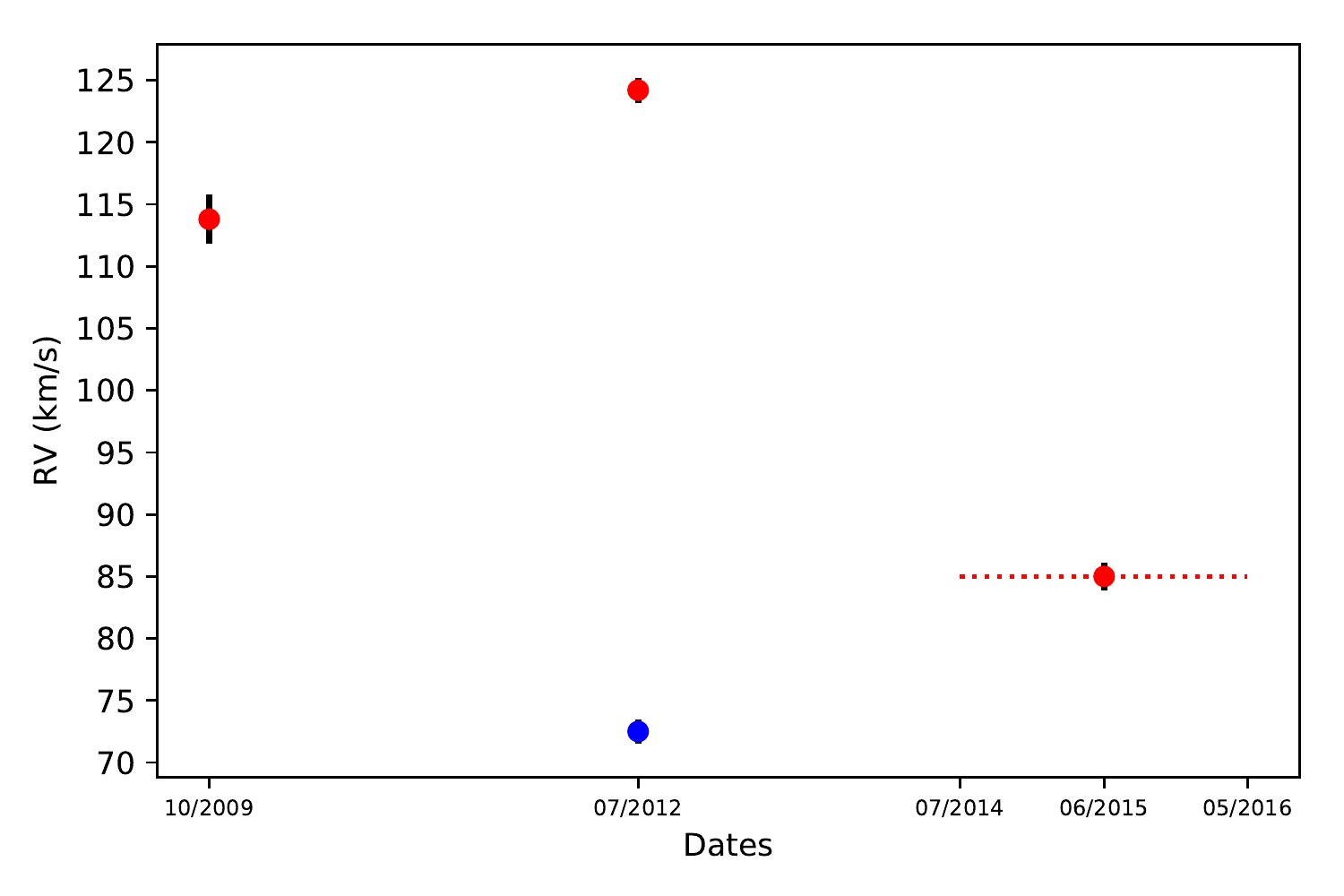}
   \includegraphics[width=\columnwidth]{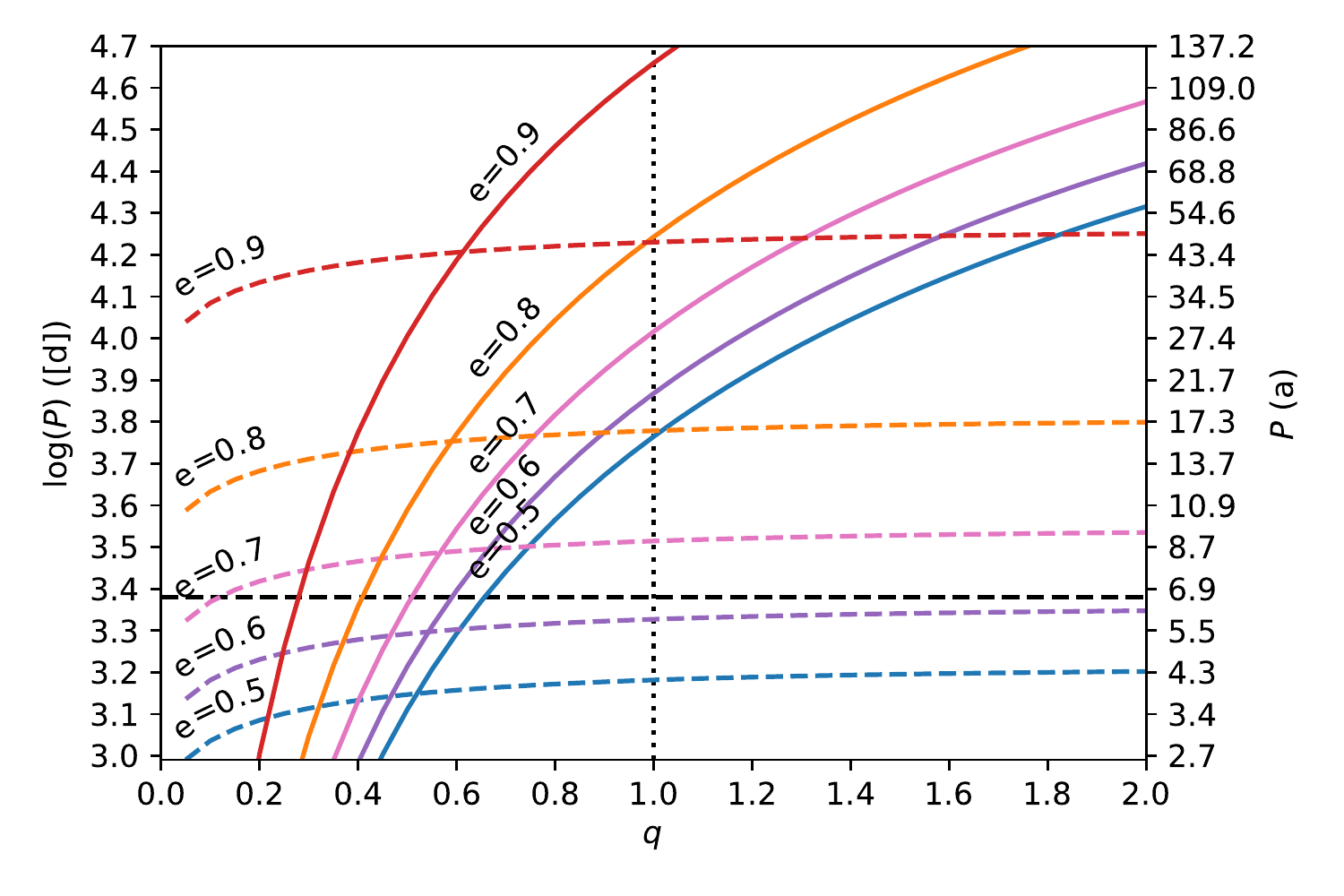} 
   \caption{
   {\bf Left (\ref{fig:ysg010}a):} RVC of YSG010. 
   Red circles are the epochs available, while the blue circle is the RV obtained for the blue companion. 
   The red dotted line show the time-span of Gaia observations.
   Whenever the error bars are not visible is because they are smaller than the circles.
   {\bf Right (\ref{fig:ysg010}b):} Study of the $P_{\rm orb}$ possible according to the estimated size of YSG010.
   Each coloured line indicates a given eccentricity.
   Dashed lines mark the $P_{\rm min}$ given its estimated size.
   Solid lines indicate the limit imposed by assuming $K_{1}=19.6$\:\kms{} (see text).
   The black-dashed horizontal line indicates $P=6.6$\:a, the minimum $P_{\rm orb}$ derived from our observations.
   The dotted vertical line indicates $q=1$.
   }
   \label{fig:ysg010}
\end{figure*}

The CSG with the largest RV variation in our sample is GDR2-ID~4685850173680692480 (YSG010 in the \citetalias{gon2015} catalogue, where it was classified as G6.5~Ia--Iab).
Despite only three epochs are available for this star within a time span of six years (see Fig.~\ref{fig:ysg010}a), it has a change in RV of $39.2$\:\kms. 
As it is relatively isolated, we discard an incorrect cross-match as an explanation for this.

YSG010 has BSSFs identified in \citetalias{gon2015}. 
We checked its spectra \citep[available in the atlas of ][]{dor2018b}, and we clearly identified the atomic lines \ion{He}{i}~4\,009\:\AA{} and 4\,026\:\AA{} from its blue companion, shifted $-52\pm1$\:\kms{} with respect to the CSG component of YSG010. 
This difference adds confidence to the $-39.2$\:\kms{} change observed.
Thus we should expect a high RV error in Gaia (see Sect.~\ref{subsub:gaia}), as YSG010 was observed five different epochs between 2014 and 2016. 
Surprisingly, it is only $1.1$\:\kms.
This indicates that its RV did not vary much during these two years, in contrast to the change of 39\:\kms{} observed between 2012 and 2014, and the difference of 52\:\kms respect its blue companion. 
The simplest explanation that reconcile these two facts is a very eccentric orbit, whose semi-major axis points roughly in our direction.
Thus, the RV change would correspond to the periapsis or to the apoapsis.
The lack of large RV variations between 2014 and 2016, together with the RV observed in 2012, indicates that the period can not be shorter than four years.
Thus implies that the RV from 2009, three years apart from the maximum in 2012, can not belong to a previous period.
In consequence, we should expect an orbital period significantly longer than the time span covered by our data (6.6\:a; from Oct. 2009 to May 2016).

We studied the minimum orbital period possible for this CSG, for a range of eccentricities (eccentricity between 0.5 and 0.9) and mass-ratios ($q$ between 0.1 and 2.0), using the methods explained in Appendix~\ref{appen:indiv_size}.
From its $M_{\rm bol}=-6.5$ and $T_{\rm eff}=4454$\:K, we estimate a mass of $\sim12$\:M$_{\odot}$ and a radius of $\sim300$\:R$_{\odot}$.
These results are shown in Figure~\ref{fig:ysg010}b.
In addition, we took into account the limits in the $P$--$q$ plane imposed by the observed change in RV : $K_{1}=39.2|/2\geq19.6$\:\kms{} for the different eccentricities considered.
As can be seen, the area of the $P$--$q$ plane where we can have $q<1$ and also satisfy our observational constraints ($P_{\rm orb}>6.6$ and at the right of the $K_{1}$ limit) is small.
From a statistical point of view, it is likely that this system has $q>1$.
In fact, if we take into account an inclination $i<90$\textdegree{} (which would shift the $K_{1}$ limit to the right), or a period a few years longer, then this system would not be compatible with $q<1$.

Although any companion whose mass was originally larger than the mass of YSG010 should have evolved (and probably died) before the former reached its current evolutionary stage, there are two scenarios able to explain a CSG in a $q>1$ system.
The first scenario is that a CSG transferred (or is transferring) a significant mass amount to the companion, but not so much that its evolution toward CSG would have been aborted.
This case can happen due to pulsation instabilities which cause a mass transfer during the periapsis in mid- to high-eccentricity systems (as is the case of well known high-mass Galactic binary systems such as VV~Cep, AZ~Cas, and KQ~Pup).
Such very specific conditions are not frequent.
According to \citet{pod1992} only 1\,--\,2\% of the high-mass stars evolve through this scenario.
The alternative scenario is a triple system.
According to \citet{moe2017}, hierarchical triple system, with a pair in a short-period orbit, and a third component orbiting the first pair, are frequent (10\%\,--\,20\%) among high-mass stars.
Thus, the triple-system scenario is one order of magnitude more likely than the scenario with a binary and mass-transferring CSG.

If YSG010 was born in a triple system, which is the most likely case, there are two possible current configurations.
In one hand, the companion is a pair of stars bonded in a short-period system, which orbits YSG010 with a very long period.
On the other hand, the original OB-type companions merged (a frequent fate among short-period OB-type stars) forming a single B-type star more massive than YSG010.
The lack of double \ion{He}{i} lines suggest that the companion is a single star, but it can be also explained if the companions have an asymmetric mass, with one of them dominating the blue flux over the other.
Thus, the available information do not allow to decide between these scenarios.
Further observations are required to confirm the $q$ of this system, and if it is $q>1$, and the nature of the companion.

%%%%%%%%%%%%%%%%%%%%%%%%%%%%%%%%%%%%%%%%%%%%%%%%%%%%%%%%%%%%%%%%%%%%%%%%%%%%
\section{Conclusions} \label{sec:conclusion}
%%%%%%%%%%%%%%%%%%%%%%%%%%%%%%%%%%%%%%%%%%%%%%%%%%%%%%%%%%%%%%%%%%%%%%%%%%%%

In this work we study the RV variations of a large sample ($\sim1\,000$) of CSGs in the MCs.
We collected data from different surveys, covering a time-span of almost 40 years and performed a statistical homogenisation to avoid systematic errors (see Sect.~\ref{subsec:homogeneisation}).
Our objective was to find CSGs in binary systems by detecting RV variations attributable to orbital motions.
For this, we studied the RV variability in our sample by two methods designed to distinguish genuine binary motions from intrinsic variations.
We found a total of 45~binary CSGs, 41 of them are previously unknown. 23 are from the SMC and 22 from the LMC (see table~\ref{table:results}).
As a result of observational biases we do no provide an estimate of the intrinsic BF but instead we calculate a lower limit of $15\pm4\%$ for the SMC and $14\pm5\%$ for the LMC.
By combining the data from both galaxies, we obtained a minimum BF of $15\pm3$\%.
These results are in in good agreement with previous studies of CSGs \citep{pat2019,pat2020} done using their RV variations, and with those obtained by \citep{neu2020} through single-epoch spectroscopy and photometric identification.

We compare our sample with the distributions obtained from OB-type stars \citep{moe2017}, which indicate that our binary population is likely dominated by low to intermediate eccentricities, intermediate mass ratios and intermediate periods.
Moreover, the comparison with the theoretical study performed by \citet{neu2018} suggest that a significant fraction of the companions in our sample are intermediate and late B-type stars. 
When we compared our results with those obtained by detecting BSSFs \citep{neu2018,neu2019}, we conclude that is vital to combine the results of both methods, BSSFs and RV variations, in order to obtain a global estimate of the BF in the future.

Finally, we report the first CSG candidate to be part of a triple system (YSG010). 
The data available suggest that is much more likely that it is (or was) a triple system, than a binary.
However, the data available is not enough to confirm this finding, and further investigation is necessary.

%%%%%%%%%%%%%%%%%%%%%%%%%%%%%%%%%%%%%%%%%%%%%%%%%%%%%%%%%%%%%%%%%%%%%%%%%%%%
\section*{Acknowledgements}
We thank Dr. Danny Lennon and Professor Ignacio Negueruela for their fruitful discussions about this topic.
We thank to the anonymous referee his/her very constructive comments.
The authors acknowledge support from the Spanish Government Ministerio de Ciencia e Innovaci\'on through grants PGC-2018-091\:3741-B-C22 and SEV 2015-0548, and from the Canarian Agency for Research, Innovation and Information Society (ACIISI), of the Canary Islands Government, and the European Regional Development Fund (ERDF), under grant with reference ProID2017010115, and the support from the Generalitat Valenciana through the grant PROMETEO/2019/041.

\section*{Data availability}
The data underlying this article are available in the article and in its online supplementary material.

%%%%%%%%%%%%%%%%%%%%%%%%%%%%%%%%%%%%%%%%%%%%%%%%%%%%%%%%%%%%%%%%%%%%%%%%%%%%
%% references
\bibliographystyle{mnras} 
\bibliography{general}

%%%%%%%%%%%%%%%%%%%%%%%%%%%%%%%%%%%%%%%%%%%%%%%%%%%%%%%%%%%%%%%%%%%%%%%%%%%%

\appendix
\section{Variation periods among CSGs}

\begin{figure}
   \centering
   \includegraphics[width=\columnwidth]{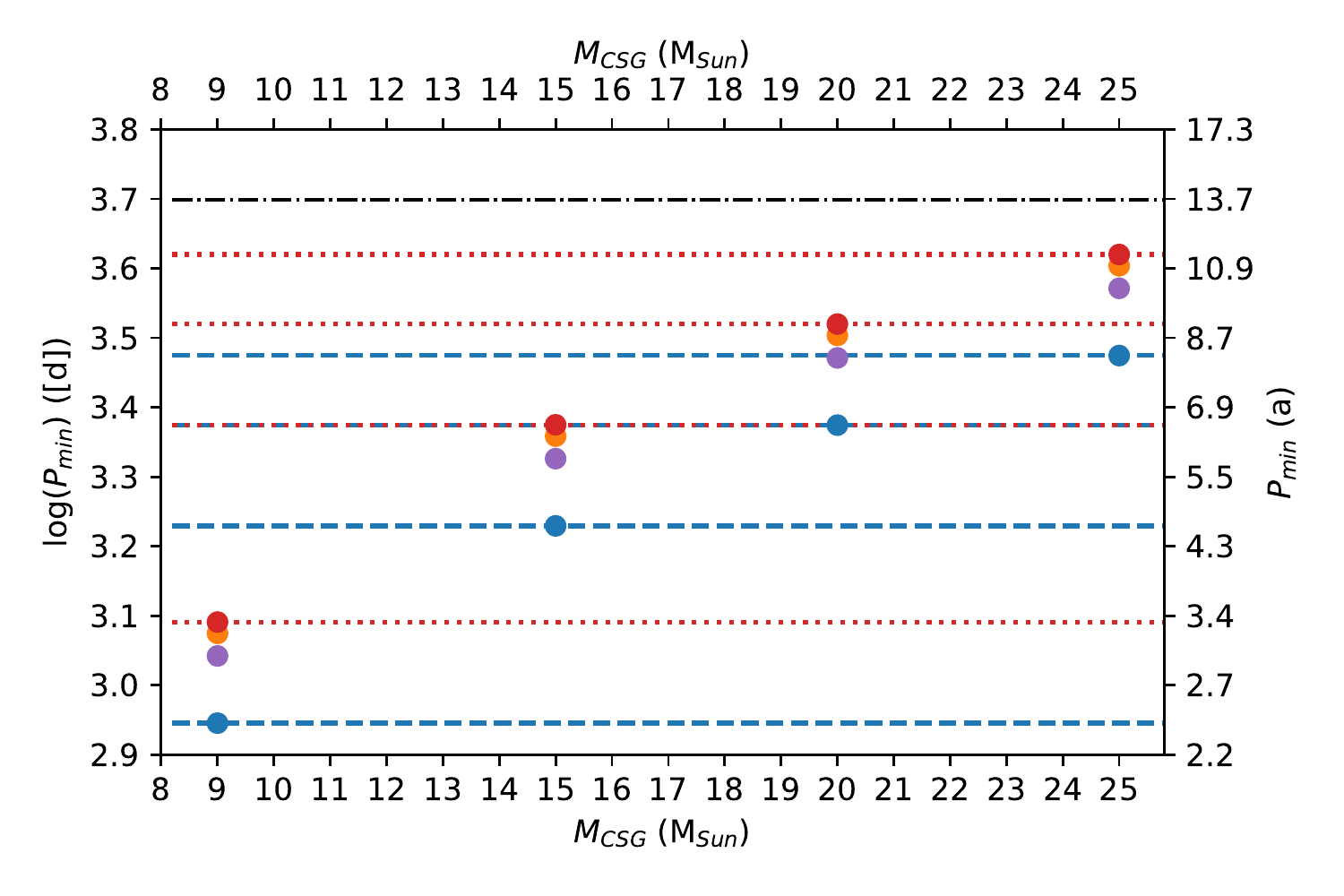}
   \includegraphics[width=\columnwidth]{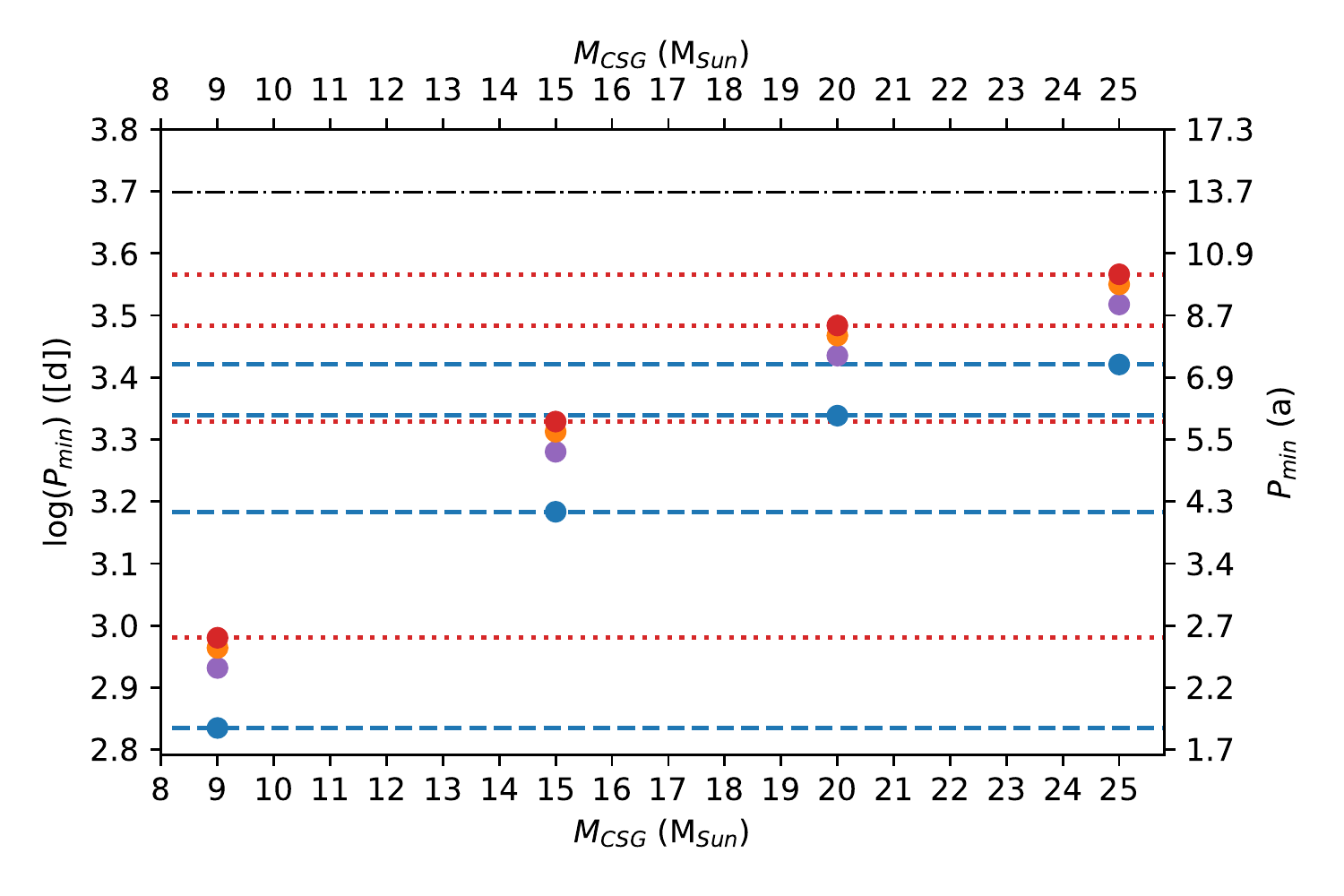}
   \caption{Minimum period depending on the CSG mass.
   The period is expressed as the $\log$ of days (left axis), and also in years (right axis).
   The colour of the circles indicate different mass ratios: blue for $q=0.1$, magenta for $q=0.4$, orange for $q=0.7$, and red for $q=1.0$.
   The black lines indicate the longest $P_{\rm 2}$ considered (5\,000\:d) for comparison. 
   The colored horizontal lines are a visual aid to better see which is the range of $P_{\rm min}$ for a given $M_{\rm CSG}$ and a different $q$.
   The upper panel show the results for CSGs from the LMC and the lower plane for CSGs from the LMC.
   }
   \label{fig:P_min}
\end{figure}

\subsection{Comparison between secondary periods and minimum orbital periods}
\label{appen:orbit}

We have calculated the minimum periods ($P_{\rm min}$) for which CSGs do not interact with their companions.
The interaction would happen if the radii of the CSG ($R_{\rm CSG}$) grow large enough to overflow its Roche Lobe at the periapsis.
By using the ratio $r_{\rm L}$ between the separation of both stars at periapsis ($r_{\rm p}$) and the size of the Roche Lobe ($r_{\rm RL}$) proposed by \citet{egg1983} it can be demonstrated that to avoid the Roche lobe overflow this condition must be satisfied:

\begin{equation*}
    R_{\rm CSG} \geq r_{\rm L} (q)\,a\,(1-e)
\end{equation*}

Where $a$ is sum of the orbital semi-major axes, and $e$ the eccentricity. 
From here we can calculate the minimum period ($P_{\rm min}$) possible for a given CSG mass ($M_{\rm CSG}$) and mass ratio $q$:

\begin{equation*}
 P_{\rm min} = \left( \frac{R_{\rm CSG}}{r_{\rm L}(q)\, (1-e)} \right)^{3/2} \, 2\, \pi\, \left( \frac{q}{G\, M_{\rm CSG}\, (1+q)} \right)^{1/2}
\end{equation*} 

We used the stellar parameters predicted by \citet{bro2011}.
In particular, we used as $R_{\rm CSG}$ the maximum size reached in any moment by the CSG of each initial mass ($M_{\rm CSG}$) considered.
The calculations were done for a number of $q$, but always with zero eccentricity, as higher eccentricity would imply necessarily longer $P_{\rm orb}$.
The results are plotted in Figure~\ref{fig:P_min}, for the SMC and the LMC respectively, and they are compared with the longest $P_{2}$ found among CSGs \citep[5\,000\:d ][]{cha2019}.

\subsection{Mass and radius estimation}
\label{appen:indiv_size}

In this paper we estimated the mass and radius for a number of CSGs.
We calculated first their luminosity using their 2MASS photometry \citep{skr2006}, transformed to the AAO system, to calculate the bolometric correction according to \citet{bes1984}.
By using infrared photometry we avoid any significant contamination from any hypothetical blue companions and we reduce the effects of interstellar reddening to a negligible level.
In addition, as the photometric variations of CSGs decrease in amplitude with the wavelength \citep{rob2008}, the infrared provides a more reliable estimation.
To calculate the absolute magnitudes, we used a distance modulus of $18.5$\:mag for the LMC \citep{wal2012} and of $19.0$\:mag for the SMC \citep{gra2014}.
Then, we estimated the mass by comparing the position of the CSG in the Hertzsprung-Russel diagram with the evolutionary tracks from \citet{bro2011}.
Finally, we calculated the radius using the luminosity and the $T_{\rm eff}$ published in \citet{tab2018} when available.
As they provide one $T_{\rm eff}$ per epoch, we use the average (weighted by their signal-to-noise ratio) as reference value.
When a CSGs is not available in that work, we used the relation between $(J-K_{\rm s})$ and $T_{\rm eff}$. 
For the LMC this relation was calculated by \citet{bri2019}.
For the SMC, we calculated this relation, and we obtained:

\begin{equation*}
    T_{\rm eff}=-1571 (J-K_{\rm s})_{0} + 5660\:({\rm K})
\end{equation*}

\section{Radial velocity curves with orbital-type behaviours}
\label{appen:curves}
In this Appendix we show the RVCs which can not be explained by intrinsic variations and thus, we have considered them caused by orbital motions (see Sect.\ref{subsec:curves}). 
We discuss each CSG in detail, and we show their corresponding RVCs.

Take into account that many of the CSGs displayed here present a high Gaia RV error. 
As explained in Sect.~\ref{subsub:gaia}, this is due these errors are related to the dispersion of the epochs measured by Gaia between 2014 and 2016. 
Thus, those with large values should not be considered as a unreliable measurement, but a sign that during these two years the RV of that CSG varied significantly. 

\begin{figure}
	\centering
	\includegraphics[width=\columnwidth]{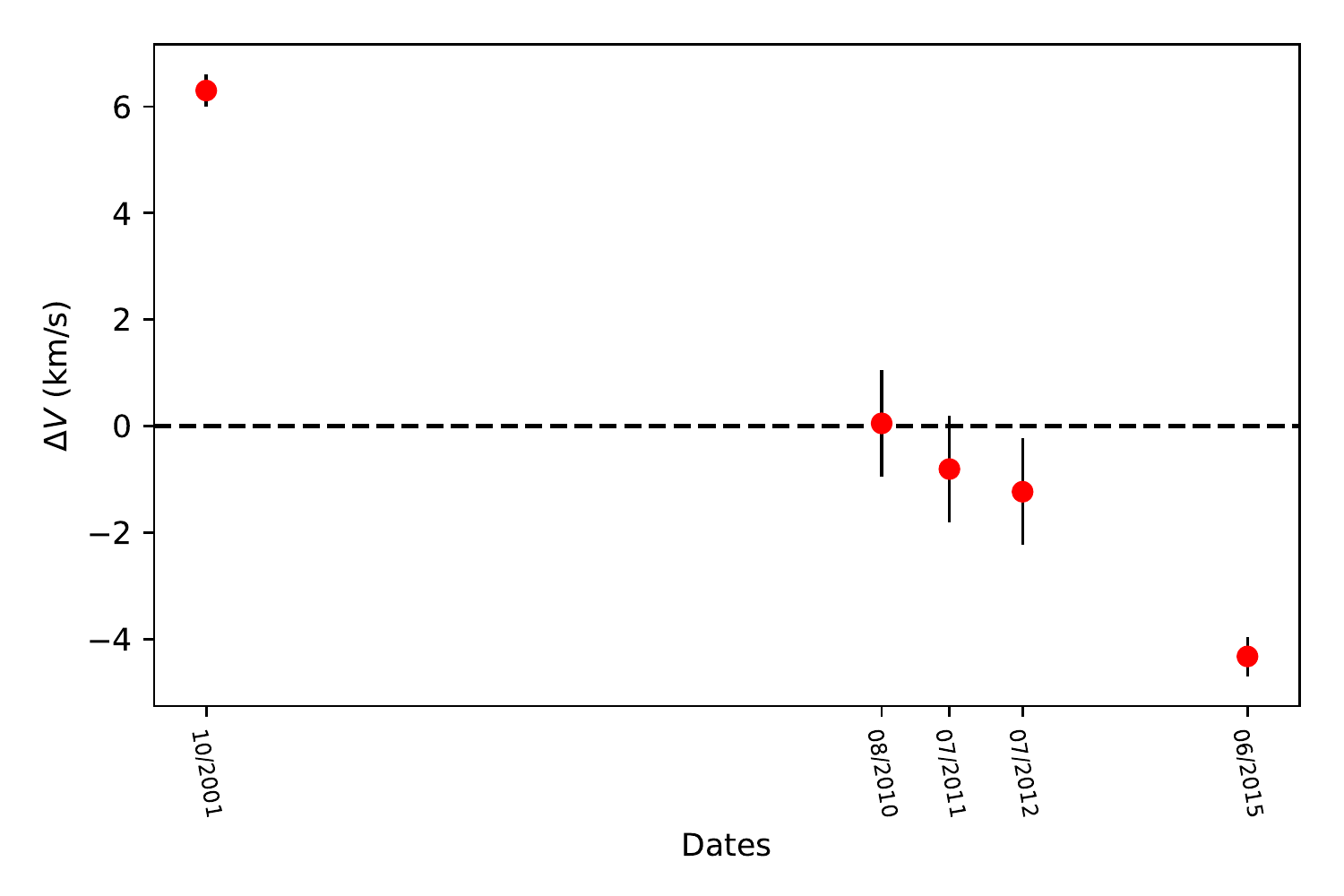}
	\caption{RVC of GDR2-ID 4685947794049430656.
	}
	\label{fig:4685947794049430656}
\end{figure}

%old B1
{\bf GDR2-ID 4685947794049430656} (Fig.~\ref{fig:4685947794049430656}): All the five data of this CSG have a Pearson's correlation coefficient $r=-0.99597$.
The probability of finding a random alignment of data points in the synthetic RVCs able to explain this coefficient is $0.01\%$, which is lower than our threshold level of $0.27\%$. 
In addition, its trend covers a variation time-span of $14$\:a, which suggest a $P_{\rm orb}$ of at least $\sim3$\:decades, although it can be much longer.
Thus, we think that intrinsic variations can not explain this curve.

\begin{figure}
	\centering
	\includegraphics[width=\columnwidth]{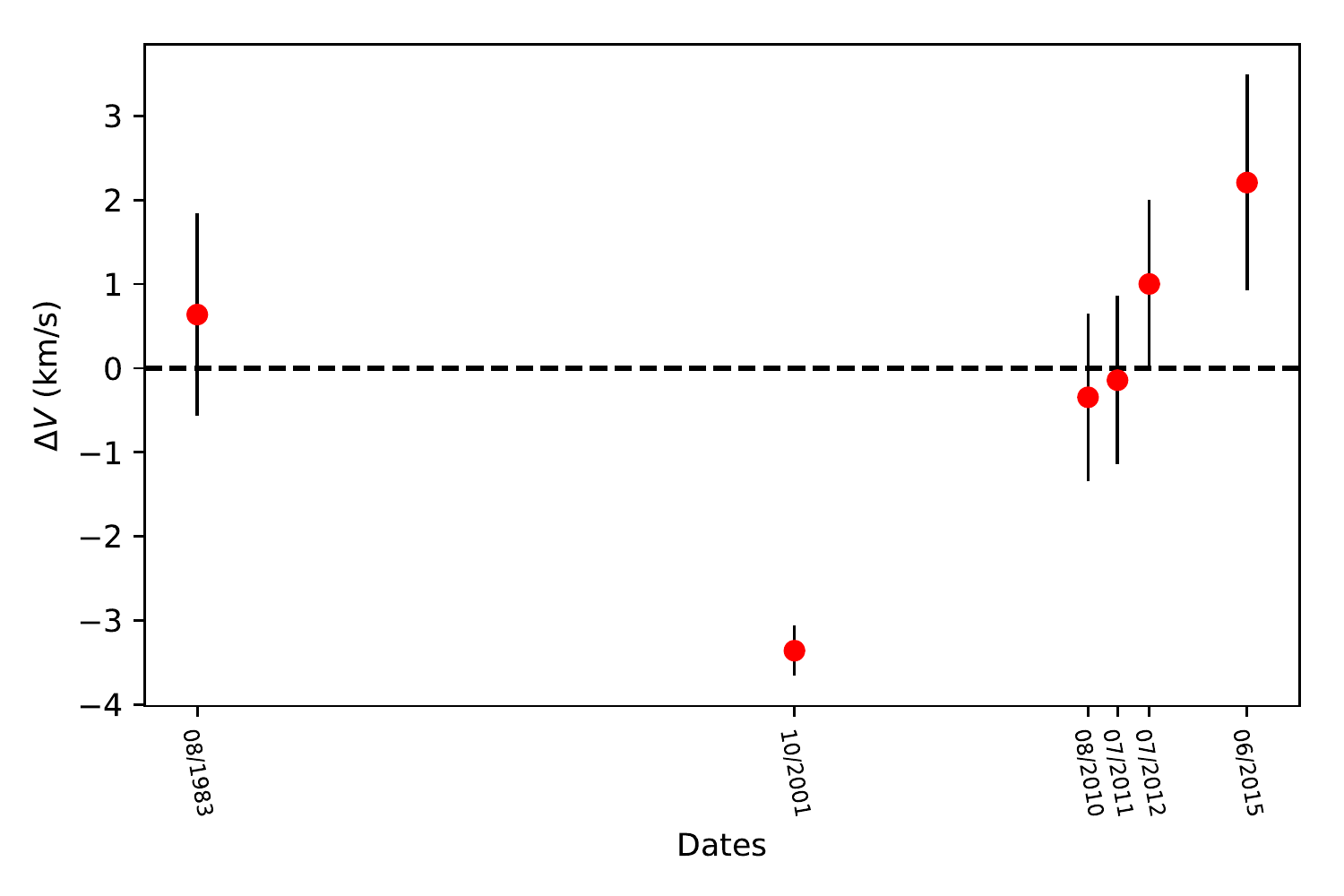}
	\caption{RVC of GDR2-ID 4690517158168014848.
	}
	\label{fig:4690517158168014848}
\end{figure}

%old B2
{\bf GDR2-ID 4690517158168014848} (Fig.~\ref{fig:4690517158168014848}): The five data between 2001 and 2015 have a Pearson's correlation coefficient $r=-0.98387$.
The probability of finding a random alignment of data points in the synthetic RVCs (with a period randomly selected from the range between 3\,928 and 4\,802\:d) able to explain this coefficient is $0.09\%$, which is lower than our threshold level of $0.27\%$.
If we assume the data points observed in 2001 and in 2015 as the minimum and maximum, respectively, of the RVC, the $P_{\rm orb}$ would be almost $30$\:a.
Thus, we think that intrinsic variations can not explain this curve.

\section{Binaries found}

\begin{table*}
\caption{List of binaries found, identified by their Gaia~DR2~ID, together with their fundamental parameters for this work.
}
\label{table:results}
\centering
\begin{tabular}{c c c c c c c c c}
\hline\hline
\noalign{\smallskip}
GDR2-ID & Catalogue name & RAJ2000 & DECJ2000 & |$\Delta V_{\rm max}$| (\kms) & $\frac{|v_{i}-v_{j}|}{\sqrt{\sigma^2_{i}+\sigma^2_{j}}}$ & \# epochs & Galaxy & Selected by\\ 
\noalign{\smallskip}
\hline
\noalign{\smallskip}
4690519185392371456 & [M2002]~50360 & 01:01:03.58 & -72:02:58.50 & 23.2 & 22.3 & 6 & SMC & |$\Delta V_{\rm max}|>11$\\ 
4690518910514527616 & [M2002]~50348 & 01:01:03.26 & -72:04:39.40 & 20.4 & 21.5 & 6 & SMC & |$\Delta V_{\rm max}|>11$\\ 
4690517158168014848 & [M2002]~48122 & 01:00:09.42 & -72:08:44.50 & 5.6 & 4.2 & 6 & SMC & RVC\\ 
4690512858880950656 & [M2002]~57386 & 01:03:47.35 & -72:01:16.00 & 31.9 & 30.5 & 4 & SMC & |$\Delta V_{\rm max}|>11$\\ 
4690509152347949184 & [M2002]~56732 & 01:03:34.30 & -72:06:05.80 & 12.1 & 11.54 & 5 & SMC & |$\Delta V_{\rm max}|>11$\\ 
4690507193841115008 & [M2002]~51000 & 01:01:19.92 & -72:05:13.10 & 14.8 & 10.5 & 6 & SMC & |$\Delta V_{\rm max}|>11$\\ 
4690505200976492160 & [M2002]~49478 & 01:00:41.56 & -72:10:37.00 & 18.1 & 7.8 & 7 & SMC & |$\Delta V_{\rm max}|>11$\\ 
4689267700627387776 & [GDN15]~SMC206 & 00:51:50.50 & -71:59:24.23 & 19.1 & 12.6 & 3 & SMC & |$\Delta V_{\rm max}|>11$\\ 
4688992032495995520 & PMMR~061 & 00:53:44.50 & -72:33:19.31 & 19.6 & 9.7 & 2 & SMC & |$\Delta V_{\rm max}|>11^a$\\ 
4688979555567809792 & [M2002]~19743 & 00:51:23.28 & -72:38:43.80 & 11.4 & 5.4 & 7 & SMC & |$\Delta V_{\rm max}|>11$\\ 
4688965850370232704 & [M2002]~18592 & 00:51:03.90 & -72:43:17.40 & 23.8 & 16.8 & 6 & SMC & |$\Delta V_{\rm max}|>11$\\ 
4688861014472093568 & SkKM13 & 00:45:04.57 & -73:05:27.51 & 15.5 & 6.6 & 6 & SMC & |$\Delta V_{\rm max}|>11$\\ 
4687501747539145216 & [GDN15]~SMC375 & 01:05:27.44 & -72:17:04.35 & 15.7 & 10.0 & 3 & SMC & |$\Delta V_{\rm max}|>11$\\ 
4687489584188959872 & [M2002]~64663 & 01:06:47.62 & -72:16:11.90 & 12.7 & 8.4 & 7 & SMC & |$\Delta V_{\rm max}|>11$\\ 
4687481024294949376 & [M2002]~55355 & 01:03:06.43 & -72:28:35.10 & 22.0 & 23.2 & 6 & SMC & |$\Delta V_{\rm max}|>11^b$\\ 
4687412064305984768 & [M2002]~63188 & 01:06:03.21 & -72:52:16.00 & 19.2 & 18.4 & 4 & SMC & |$\Delta V_{\rm max}|>11$\\ 
4686455764056822016 & [M2002]~83593 & 01:30:33.92 & -73:18:41.90 & 12.2 & 7.4 & 3 & SMC & |$\Delta V_{\rm max}|>11$\\ 
4686417315506422016 & PMMR~198 & 01:18:17.70 & -73:09:27.48 & 13.9 & 5.6 & 3 & SMC & |$\Delta V_{\rm max}|>11$\\ 
4685989781659094144 & [GDN15]~YSG070 & 01:00:09.60 & -72:33:59.30 & 16.2 & 7.1 & 3 & SMC & |$\Delta V_{\rm max}|>11$\\ 
4685947794049430656 & [M2002]~11101 & 00:48:31.92 & -73:07:44.40 & 10.6 & 22.4 & 5 & SMC & RVC\\ 
4685926761539086848 & [M2002]~11939 & 00:48:51.82 & -73:22:39.83 & 12.1 & 6.1 & 6 & SMC & |$\Delta V_{\rm max}|>11$\\ 
4685850173680692480 & [GDN15]~YSG010 & 00:47:08.69 & -73:14:11.90 & 39.2 & 25.9 & 3 & SMC & |$\Delta V_{\rm max}|>11^b$\\ 
4685836605893771520 & [GDN15]~SMC099 & 00:46:41.68 & -73:22:54.20 & 16.0 & 6.9 & 2 & SMC & |$\Delta V_{\rm max}|>11^b$\\ 
4662153228519937664 & RM1-080 & 04:57:26.36 & -66:23:25.78 & 13.06 & 4.7 & 3 & LMC & |$\Delta V_{\rm max}|>11$\\ 
4661877491594670208 & RM1-151 & 05:04:14.12 & -67:16:14.40 & 13.84 & 4.8 & 3 & LMC & |$\Delta V_{\rm max}|>11$\\ 
4661772144640288384 & RM1-024 & 04:52:53.64 & -66:55:52.14 & 11.93 & 9.6 & 3 & LMC & |$\Delta V_{\rm max}|>11$\\ 
4661446757944600448 & 2MASS~J05045392-6801585 & 05:04:53.93 & -68:01:58.64 & 24.14 & 4.5 & 2 & LMC & |$\Delta V_{\rm max}|>11$\\
4661306016119437056 & 2MASS~J05030232-6847203 & 05:03:02.35 & -68:47:20.23 & 29.94 & 6.0 & 2 & LMC & |$\Delta V_{\rm max}|>11^a$\\
4660179639518872576 & [M2002]~147928 & 05:30:33.47 & -67:17:15.07 & 23.47 & 14.4 & 3 & LMC & |$\Delta V_{\rm max}|>11$\\ 
4658466355538857728 & RM2-093 & 05:31:24.27 & -68:41:33.64 & 14.25 & 6.2 & 4 & LMC & |$\Delta V_{\rm max}|>11^b$\\ 
4658433443240057216 & [M2002]~145716 & 05:29:54.75 & -69:04:15.69 & 12.52 & 7.1 & 7 & LMC & |$\Delta V_{\rm max}|>11$\\ 
4658429143945343872 & [M2002]~150577 & 05:31:18.44 & -69:09:28.16 & 25.29 & 4.6 & 6 & LMC & |$\Delta V_{\rm max}|>11$\\ 
4658287100783922304 & 2MASS~J05191579-6856039 & 05:19:15.79 & -68:56:03.89 & 18.47 & 5.1 & 2 & LMC & |$\Delta V_{\rm max}|>11$\\
4658097198899548928 & [M2002]~139413 & 05:27:47.51 & -69:13:20.59 & 11.73 & 8.3 & 6 & LMC & |$\Delta V_{\rm max}|>11^a$\\ 
4658057685185383424 & [M2002]~144217 & 05:29:27.58 & -69:08:50.33 & 11.69 & 4.5 & 5 & LMC & |$\Delta V_{\rm max}|>11$\\ 
4658052943542064896 & [M2002]~148381 & 05:30:41.43 & -69:15:33.84 & 14.06 & 13.5 & 4 & LMC & |$\Delta V_{\rm max}|>11^b$\\
4657674642790217600 & [M2002]~174742 & 05:40:25.32 & -69:15:30.20 & 11.46 & 5.1 & 5 & LMC & |$\Delta V_{\rm max}|>11$\\ 
4657598364163823872 & SP77~54-38 & 05:41:10.66 & -69:38:04.04 & 11.94 & 7.3 & 4 & LMC & |$\Delta V_{\rm max}|>11$\\ 
4657099387712019072 & RM1-603 & 05:33:04.42 & -70:48:30.13 & 16.14 & 14.0 & 3 & LMC & |$\Delta V_{\rm max}|>11$\\ 
4655456756015488000 & [M2002]~21369 & 04:54:36.85 & -69:20:22.18 & 12.34 & 7.1 & 5 & LMC & |$\Delta V_{\rm max}|>11$\\ 
4655369100006728320 & RM1-015 & 04:51:20.58 & -69:29:13.97 & 16.82 & 4.7 & 3 & LMC & |$\Delta V_{\rm max}|>11$\\ 
4655321000684432384 & 2MASS~J04510944-6956146 & 04:51:09.48 & -69:56:14.43 & 14.4 & 4.7 & 2 & LMC & |$\Delta V_{\rm max}|>11$\\ 
4651906394290251520 & RM2-069 & 05:23:35.76 & -70:44:55.34 & 15.38 & 3.1 & 3 & LMC & |$\Delta V_{\rm max}|>11$\\ 
4651885194331239424 & 2MASS~J05210134-7058459 & 05:21:01.36 & -70:58:45.90 & 16.61 & 4.8 & 2 & LMC & |$\Delta V_{\rm max}|>11$\\
4657696972286770432 & [M2002]~164506 & 05:36:06.37 & -68:56:40.70 & 11.30 & 8.4 & 5 & LMC & |$\Delta V_{\rm max}|>11$\\ 
\noalign{\smallskip}
\hline
\multicolumn{9}{l}{\footnotesize $^a$ This CSG has been reported to have BSSFs in \citet{neu2019}}\\
\multicolumn{9}{l}{\footnotesize $^b$ This CSG has been reported to have BSSFs in \citet{gon2015}}\\
\end{tabular}
\end{table*}

%%%%%%%%%%%%%%%%%%%%%%%%%%%%%%%%%%%%%%%%%%%%%%%%%%

% Don't change these lines
\bsp	% typesetting comment
\label{lastpage}
\end{document}